\documentclass[journal]{IEEEtran}
\usepackage{amssymb}
\usepackage{amsfonts}
\usepackage{amsfonts,subfigure,multicol,color,verbatim,
graphicx,cite,epsfig,amssymb,amsmath,cases,bm,algorithm,
algorithmic,xcolor,multirow,array}

\begin{document}
\markboth{IEEE Communications Surveys \& Tutorials, Vol. X, No. Y,
Mon. 2016} {Peng: C-RANs \ldots}

\title{Recent Advances in Cloud Radio Access Networks: System Architectures, Key Techniques, and Open Issues}

\author{\normalsize
Mugen~Peng, Yaohua~Sun, Xuelong Li, Zhendong~Mao, and Chonggang~Wang
\thanks{Manuscript received Sep. 24, 2015; First Revised Dec. 23, 2015; Second Revised Feb. 04, 2016; Accepted Mar. 19, 2016.}
\thanks{This work was supported in part by the National Natural Science
Foundation of China under Grant No. 61361166005, the
National High Technology Research and Development Program of China
under Grant No. 2014AA01A701, the National Basic Research Program of
China (973 Program) under Grant No. 2013CB336600, and the State Major Science and Technology Special Projects (Grant No. 2016ZX03001020-006).}
\thanks{Mugen~Peng (e-mail: {\tt pmg@bupt.edu.cn}), Yaohua~Sun (e-mail: {\tt sunyaohua@bupt.edu.cn}),
and Zhendong~Mao (e-mail: {\tt mzd@bupt.edu.cn}) are with the Key Laboratory of Universal Wireless Communications for
Ministry of Education, Beijing University of Posts and
Telecommunications, China. Xuelong Li (e-mail: {\tt xuelong\_li@opt.ac.cn}) is with the Center for OPTical IMagery Analysis and Learning (OPTIMAL), State Key Laboratory of Transient Optics and Photonics, Xi'an Institute of Optics and Precision Mechanics, Chinese Academy of Sciences, Xi'an 710119, Shaanxi, P. R. China. Chonggang~Wang (e-mail: cgwang@ieee.org)
is with the InterDigital Communications, King of Prussia, PA, USA.
}}

\maketitle

\begin{abstract}
As a promising paradigm to reduce both capital and operating expenditures, the cloud radio access network (C-RAN) has been shown to provide high spectral efficiency and energy efficiency. Motivated by its significant theoretical performance gains and potential advantages, C-RANs have been advocated by both the industry and research community. This paper comprehensively surveys the recent advances of C-RANs, including system architectures, key techniques, and open issues. The system architectures with different functional splits and the corresponding characteristics are comprehensively summarized and discussed. The state-of-the-art key techniques in C-RANs are classified as: the fronthaul compression, large-scale collaborative processing, and channel estimation in the physical layer; and the radio resource allocation and optimization in the upper layer. Additionally, given the extensiveness of the research area, open issues and challenges are presented to spur future investigations, in which the involvement of edge cache, big data mining, social-aware device-to-device, cognitive radio, software defined network, and physical layer security for C-RANs are discussed, and the progress of testbed development and trial test are introduced as well.
\end{abstract}


\section{Introduction}

Mobile data traffic has been growing exponentially over the past few years. A report from Cisco shows that the mobile data traffic in 2014 grew 69 percent and was nearly 30 times the size of the entire global Internet in 2000 \cite{t-1}. It is predicted that the global mobile data traffic will increase nearly tenfold between 2014 and 2019 \cite{II:IBM}. One of the primary contributors to the explosive mobile traffic growth is the rapid proliferation of mobile social applications running on mobile smart devices, and these sharp increases in mobile traffic (particularly from bandwidth-hungry mobile applications) are projected to continue in the foreseeable future. To combat the rapidly growing demand, the next-generation wireless access networks are expected to offer high spectral efficiency (SE). On the other hand, the rapid growth of operating expenditure and energy consumption has becomes problematic, especially with the slow increase of mobile operators' profit \cite{II:Mobile}. In 2012, the annual average power consumption by information technology (IT) and communication industries was over 200 gigawatt, of which about 25\% was from the telecommunication infrastructure and mobile devices \cite{zte}.

The wireless world is increasingly calling for advanced intelligent wireless network architectures to reduce networking cost and meet the future requirements of high SE and low power consumption. Since most energy is consumed by base stations (BSs), it is appealing to move the functions of BSs into the centralized cloud server so as to optimize resource and energy consumption \cite{intel,bell}. Accordingly, the cloud radio access network (C-RAN) has recently been recognized as an evolved system paradigm by both the operators and the equipment vendors that curtails both capital and operating expenditures, while providing high energy-efficiency transmission bit rates in the advanced wireless communication systems \cite{ita,do,nokia}.

\begin{figure}[!h]
\centering \vspace*{0pt}
\includegraphics[scale=0.5]{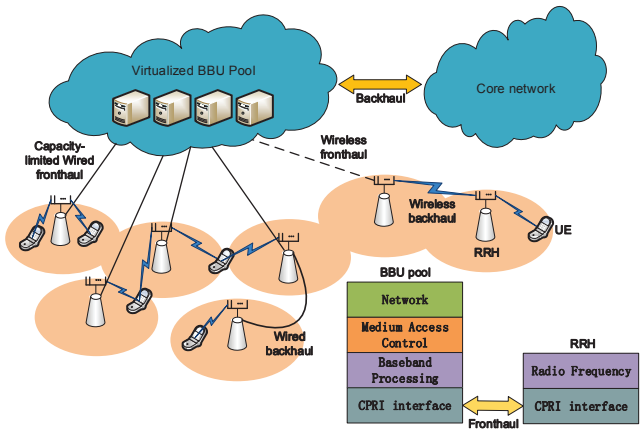}
\setlength{\belowcaptionskip}{-100pt} \vspace*{-10pt}
\caption{A general C-RAN system}\label{c-ran}
\end{figure}

The C-RAN is a breakthrough combination of emerging technologies from both the wireless network and the information technology industries that has the potential to improve SE and energy efficiency (EE) based on the centralized cloud principle of sharing storage and computing resources via virtualization. In a general C-RAN as shown in Fig. \ref{c-ran}, the traditional BS is decoupled into two parts: distributed remote radio heads (RRHs), and baseband units (BBUs) clustered as a BBU pool in a centralized location. RRHs with radio frequency functions support high capacity in hot spots, while the virtualized BBU pool provides large-scale collaborative processing (LSCP), cooperative radio resource allocation (CRRA), and intelligent networking. The BBU pool communicates
with RRHs via common public radio interface (CPRI) protocol, which supports a constant bit rate and bidirectional digitized in-phase and quadrature (I/Q) transmission, and includes specifications for control plane and data plane \cite{cpri}.

The wireless industry and research community are working on the vision and requirements of C-RANs, including the reduced cost, high EE and SE, forward and backward compatibility. In particular, C-RANs can reduce the cost of the deployment owing to the possibility to substitute full-fledged dense BSs with RRHs. The RRH deployment results in reducing the construction space and saving energy consumptions of dense BSs. In \cite{cost}, a quantitative analysis of the cost was presented in C-RANs, and it was shown that C-RANs can lead to a capital expenditure reduction with $10\%$ to $15\%$ per kilometer comparing with traditional long term evolution (LTE) networks. Meanwhile, C-RANs enable the flexible allocation of scarce radio and computing resources across all RRHs centrally processed by the same BBU pool, which reaps statistical multiplexing gains due to the load balance of adjacent RRHs for the un-uniformly traffic distribution. In \cite{scp}, the equal per-cell rate was analyzed for half-duplex and full-duplex operations in C-RANs, and the simulation results demonstrated that significant performance gains can be achieved by both operation modes compared with the corresponding single-cell processing approaches under unlimited fronthaul capacity. Furthermore, C-RANs simplify the cellular network upgrading and maintenance due to the centralized architecture. At the BBU pool, baseband processing can be implemented using software radio technology based on open IT architectures, making the systems¡¯ upgrading to different standards possible without hardware upgrades. With C-RANs, mobile operators can quickly deploy RRHs to expand and make upgrades to their cellular network. Mobile operators only need to install new RRHs and connect them to the BBU pool in order to expand the network coverage or split the cell to improve SE.

Motivated by the potential significant benefits, C-RANs have been advocated by both mobile operators (e.g., France Telecom/Orange, NTT DoCoMo, Telefonica, and China Mobile) and equipment vendors (e.g., Cisco, Intel, IBM, Huawei, ZTE, Ericsson, and so on). Despite the mentioned attractive features, C-RANs also come with their own technological challenges. For example, deploying a large number of RRHs entails significant interference when LSCP works inefficiently due to the constrained fronthaul, which presents a bottleneck to the capacity of C-RANs. In practice, in order to mitigate interference, it is necessary to exploit the large-scale precoding and decoding schemes at the BBU pool that takes into account the fronthaul capacity limitations, implementation complexity, and imperfections in the channel state information (CSIs) collected at the BBU pool from all connected RRHs. Particularly, in the uplink, each RRH compresses its received signal for transmission to the central BBU pool via the constrained fronthaul link. The central BBU pool then performs joint decoding of the data streams of all user equipments (UEs), based on the received compressed signals. Compression over the constrained fronthaul and large-scale decoding pose significant technical challenges to be tackled for a successful rollout and commercial operations. In the downlink, each UE receives signals from adjacent RRHs and interfering RRHs, which include interference and useful signal. Precoding at the BBU pool, along with channel estimation and detection at the UE, pose the technical challenges. Additionally, radio resource management is a key functionality run at the BBU pool to decrease the interference and achieve the potential performance gains.

Recently, diverse problems and corresponding solutions for C-RANs have been intensively elaborated from
the aspects of system architecture, fronthaul compression, LSCP, and CRRA, allowing feasible designs and operations of energy efficient and
spectrum efficient C-RANs. To systematically show the advances in C-RANs, several survey works have been done \cite{s1,s2,s3,s4,front10,r8}. The progress on the centralization and virtualization for C-RANs is introduced in \cite{s1}, where the performances of several key technologies like uplink coordinated multiple point (CoMP) are verified, and a general purpose platform (GPP) based C-RAN test bed is built, showing almost the same performance as
traditional systems. While in \cite{s2}, a high-level overview of C-RANs
is provided, including the architecture, transport network techniques, virtualization and challenges.
A novel logical structure of
C-RANs comprising physical plane, control plane, and service plane is presented in \cite{s3}, along with
a coordinated user scheduling algorithm and parallel optimum precoding scheme.
In \cite{s4}, a comprehensive survey related to recent advances in fronthaul-constrained C-RANs is discussed, mainly focusing on
potential solutions to enhance SE and EE.
In addition, the authors in \cite{front10} reviewed
the key fronthaul compression techniques for the uplink and downlink of a C-RAN, together with simulation results to validate the performance improvement
desired from the implementation of multiterminal fronthaul compression. In \cite{r8}, the developments of resource allocation in heterogeneous cloud radio access networks (H-CRANs) are investigated, and three resource allocation schemes are proposed, i.e., coordinated scheduling, hybrid backhauling, and multicloud association.

Although initial efforts have been made on the survey of C-RANs, limitations exist in the previous works that the content is only related to the practical implementation and deployment aspects in \cite{s1}, or only a small part of current research achievements on C-RANs are specified in \cite{s3}, or only a specific topic, which is relative to H-CRANs but not the general C-RANs, is concentrated on in \cite{r8}. Considering these shortcomings and the ongoing research activities, a more
comprehensive survey framework for incorporating the basics and the latest achievements on the system architecture and key techniques across different layers over the air interface seems timely and significant. As shown in Table \ref{contributions}, compared with the existing survey articles, a
more comprehensive survey framework for recent advances in system architectures and key techniques to improve SE and EE is proposed in this paper.
Specifically, this envisioned paper surveys the state-of-the-art system architectures, key technologies on the advanced LSCP, fronthaul compression, and channel estimation in PHY, as well as CRRA in the upper layers. Additionally, given the extensiveness of the research area, open issues and challenges are presented to spur future investigations. The aims of this survey are to fill the research gaps found in the previous publications \cite{s1,s2,s3,s4,front10,r8} as stated above. In this way, the contributions of this paper are fourfold as summarized below:

\begin{enumerate}
 \item A comprehensive survey of system architectures is presented, which is divided into the architectures proposed by the industry and the
 architectures proposed by the academia. Particularly, architectures proposed by the industry are illustrated based on different functional splits, in which the tradeoff between implementation complexity and performance gains is concerned. The system architecture evolution to H-CRANs and fog computing based radio access networks (F-RANs) is highlighted in the research community.
 \item A comprehensive survey of key techniques in PHY is presented, including the fronthaul compression, LSCP, and channel estimation. The principle, challenges and the technical solutions of these key techniques in PHY are summarized, and the contributions and researching results of relative literatures are systematically elaborated. Particularly, the quantization, compressive sensing (CS), and spatial filtering for the fronthaul compression in both uplink and downlink are summarized. Meanwhile, both the linear LSCP with/without perfect CSIs and the nonlinear sparse LSCP are presented. In addition, the superimposed training, segment training, and semi-blind channel estimation approaches for C-RANs are surveyed.
 \item The CRRA for C-RANs are comprehensively summarized, including the static CRRA without considering queue state information (QSI) and the dynamic CRRA with queue-awareness. In particular, the optimization approaches for the static CRRA are classified into classic non-convex optimization approaches and game model based approaches. The dynamic CRRA is surveyed into the equivalent rate approach, the Lyapunov optimization approach, and the Markov decision process approach.
 \item The future challenges and open issues related to C-RANs are identified, as
edge cache, big data mining, social-aware device-to-device (D2D) communication, cognitive radio (CR), software defined network (SDN), physical layer security, and trial test are vital for the further development of C-RANs.
\end{enumerate}

\begin{table}[h]
\caption{Summary of Existing Survey Articles}\label{contributions}
\scriptsize 
\centering
\begin{tabular}{|p{0.7 in}|p{0.3 in}|p{2 in}|}
 \hline
 \textbf{Aspects} & \textbf{Survey Papers}& \textbf{Contributions} \\ \hline
 \multirow{5}{0.7 in}{Comprehensive Surveys}
 & \cite{s1} & Initial results from the field trials of the centralization and achievements on the prototype development of virtualized C-RANs. \\ \cline{2-3}
 & \cite{s2} & A high level overview of traditional C-RANs including the architecture, the transport network, the virtualization, and so on. \\ \cline{2-3}
 & \cite{s3} & A logical structure of C-RANs comprising physical plane, control plane, and service plane. \\ \cline{2-3}
 & \cite{s4} & Recent advances in fronthaul-constrained C-RANs, focusing on potential solutions to enhancing SE and EE. \\ \cline{2-3}
 & This Article & A comprehensive survey of state-of-the-art system architectures, key technologies on the advanced LSCP in PHY, and CRRA in the upper layers. \\ \hline
 \multirow{2}{0.7 in}{Fronthaul Compression and Quantization}
 & \cite{front10} & A review of quantization based fronthaul compression techniques for both the uplink and downlink of a C-RAN.\\ \cline{2-3}
 & This Article & A comprehensive survey of latest fronthaul compression techniques, including quantization based fronthaul compression, compressive sensing based compression, and spatial filtering.\\ \hline
 Large-Scale Collaborative Processing &This Article & A comprehensive survey of precoding techniques with full CSIs, precoding techniques with partial CSIs, and sparse precoding.\\ \hline
 Channel Estimation and Training Design &This Article & An overview of channel estimation techniques recently proposed for C-RANs, giving useful guidelines on the new requirements when employing the conventional channel estimation approaches in C-RANs. \\ \hline
 \multirow{2}{0.7 in}{Radio Resource Allocation and Optimization}
 & \cite{r8} & An investigation of developments in centralized approaches to resource allocation only in the special H-CRANs but not for a general C-RAN.\\ \cline{2-3}
 & This Article & A full survey of optimization approaches to static CRRA without considering QSI and dynamic CRRA with queue-awareness in C-RANs, covering both classic approaches and game model based approaches.\\ \hline
\end{tabular}
\end{table}

The rest of this paper is organized as follows: Section II surveys the system architectures
of C-RANs from the literatures related to the industry and research community. The uplink and downlink compression to decrease the capacity requirements of fronthaul are summarized in Section III. In Section IV, the LSCP techniques for C-RANs
are surveyed. Section V summarizes recent advances of channel estimation and training design solutions. The static and dynamic CRRA are surveyed in VI. The current challenges and open
issues are shown in Section VII, prior to the conclusion in Section VIII. For convenience, all abbreviations are listed in Table II.

\begin{table}\label{table2}
\center \caption{SUMMARY OF ABBREVIATIONS}
\begin{tabular}{l l}\hline
 3G & $3^{rd}$ generation  \\
 4G & $4^{th}$ generation  \\
 5G & $5^{th}$ generation  \\
 ACL & radio access link \\
 BBU & baseband unit \\
 BS & base station  \\
 CA & carrier aggregation  \\
 CoMP & coordinated multiple point  \\
 CPRI & common public radio interface  \\
 CR & cognitive radio  \\
 C-RAN & cloud radio access network \\
 CRRA & cooperative radio resource allocation  \\
 CS & compressive sensing \\
 CSI & channel state information \\
 CWDM & coarse wavelength division multiplexing \\
 D2D & device-to-device  \\
 D-RAN & distributed radio access network  \\
 DAS & distributed antenna system  \\
 DPR & decompress-process-and-recompress  \\
 DWDM & dense wavelength division multiplexing  \\
 ECF & estimate-compress-forward  \\
 EE & energy efficiency  \\
 EPON & Ethernet PON \\
 F-RAN & fog computing based radio access network \\
 GSBF & group sparse beamforming \\
 GSM & global system for mobile communications \\
 H-CRAN & heterogeneous cloud radio access network \\
 I/Q & in-phase and quadrature \\
 LMMSE & linear minimum-mean-square-error  \\
 LS & least-squared  \\
 LSCP & large-scale collaborative processing  \\
 LTE & long term evolution  \\
 MA & multiple access  \\
 MAC & medium access control \\
 MARN & multiple access relay networks\\
 MBS & macro BS \\
 MDP & Markov decision process \\
 MF & multiplex-and-forward \\
 NG-PON & next-generation passive optical network \\
 NGMN & next generation mobile networks \\
 MIMO & multiple input multiple output \\
 ML & maximum likelihood \\
 MU-MIMO & multiuser MIMO \\
 PE & polynomial expansion \\
 PHY & physical \\
 QoE & quality of experience \\
 QoS & quality of service \\
 QSI & queue state information \\
 RANaaS & radio access network as a service \\
 RF & radio frequency \\
 RRH & remote radio head \\
 RRM & radio resource management\\
 SDF &software defined fronthaul \\
 SDN &software defined network \\
 SE & spectrum efficiency \\
 SINR & signal-to-interference-plus-noise ratio \\
 SQNR & signal-to-quantization-noise-ratio \\
 SS & superimposed-segment \\
 TD-LTE & time division-long term evolution \\
 TD-SCDMA & time division-synchronization code \\
   &       division multiple access \\
 TDMA & time division multiple access \\
 TWDM & time wavelength division multiplexing \\
 UE & user equipment\\
 WDM & wavelength-division multiplexing \\
 WSR & weighted sum rate \\
 WMMSE & weighted minimum mean square error \\
 WNC & wireless network cloud \\ \hline

\end{tabular}
\end{table}

\section{System Architectures}

In recent years, C-RANs have attracted much attention of both industries and the academia due to their significant benefits to
meet the enormous demand for data traffic and meanwhile decrease the capital and operational expenditures.
In the industry, many operators and equipment vendors have actively expressed their own opinions on C-RAN architectures in the form of technical reports or
literatures. While in the academia, different solutions have been proposed for the enhancements of C-RANs.
In this section, the general system architecture of C-RANs is firstly introduced, and then the specific architectures in the industry and academia are briefly surveyed.

\subsection{C-RAN General Architectures}

The evolution of cellular BSs in the mobile communication system is shown in Fig. \ref{revolution}.
In 1G and 2G cellular networks, radio and baseband processing
functionalities are integrated inside a BS. While in 3G and 4G cellular networks,
a BS is divided into a RRH and a BBU.
In this BS architecture, the location of the BBU can be far from the RRH
for a low site rental and the convenience of maintenance.
In the 4G beyond, BBUs are migrated into a BBU pool, which
is virtualized and shared by different cell sites, and RRHs are connected to the BBU pool via
the fronthaul links, which results in a general C-RAN architecture.


By the centralized LSCP in C-RANs, several benefits can be achieved.
First, a huge energy consumption incurred by air conditioning
can be saved owing to the reduction of BS sites.
Second, the operation and maintenance cost can be decreased because BBUs are placed in several big rooms. Third, the LSCP technique
can be implemented more easily due to the centralized structure than the traditional cellular system, which significantly improves SE.
Fourth, C-RANs can be endowed with the capability of load balancing in the BBU pool, and thus adapts to the non-uniform traffic.

\subsection{C-RAN Architectures Proposed by Industries}

\begin{figure}[!h]
\centering \vspace*{0pt}
\includegraphics[scale=0.4]{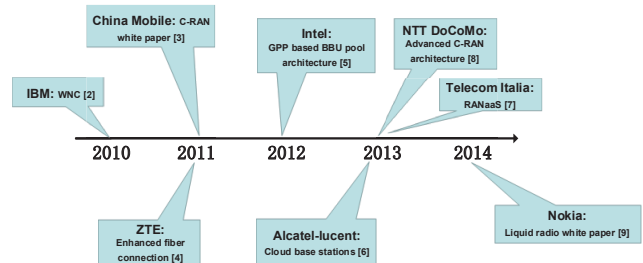}
\setlength{\belowcaptionskip}{-100pt} \vspace*{-10pt}
\caption{The milestone of C-RAN architectures in the industry.}\label{mile}
\end{figure}

As shown in Fig. \ref{mile}, the history of C-RANs in the industry can be traced back to 2010, in which the concept of C-RANs is firstly proposed with the name of wireless network cloud (WNC) by IBM to
decrease networking cost and obtain more flexible network capabilities \cite{II:IBM}. Then this concept is further
exploited by China Mobile Research Institute in 2011 \cite{II:Mobile}, in which
the C-RAN architecture is elaborated along with the technology trends and feasibility analysis.
To deal with the fiber scarcity, ZTE proposed different solutions, including enhanced fiber
connection, colored fiber connection, and optical transport network bearer,
which can use a very few fibers to meet the bearer requirements of C-RANs\cite{zte}.
An efficient and scalable GPP based BBU pool architecture is proposed by Intel,
and it can utilize computation resource as needed \cite{intel}.
Leveraging virtualization techniques, the concept of cloud BSs
is proposed by Alcatel-Lucent, which requires less
processing resources without degrading system performances \cite{bell}.
In 2013, NTT DoCoMo began developing
the advanced centralized C-RAN architecture for its future LTE-Advanced mobile system \cite{do}.
Also in 2013, the RAN-as-a-Service (RANaaS) concept is emphasized by Telecom Italia,
which provides the benefits of the C-RAN architecture with more flexibility \cite{ita}.
While in 2014, the white paper of Liquid Radio is released by Nokia Networks, in which the centralized C-RAN is taken as an efficient way
to enhance the network utilization \cite{nokia}.

\subsubsection{\textbf{Functional Split of C-RANs}}

\begin{figure*}[!htp]
\centering \vspace*{0pt}
\includegraphics[scale=0.55]{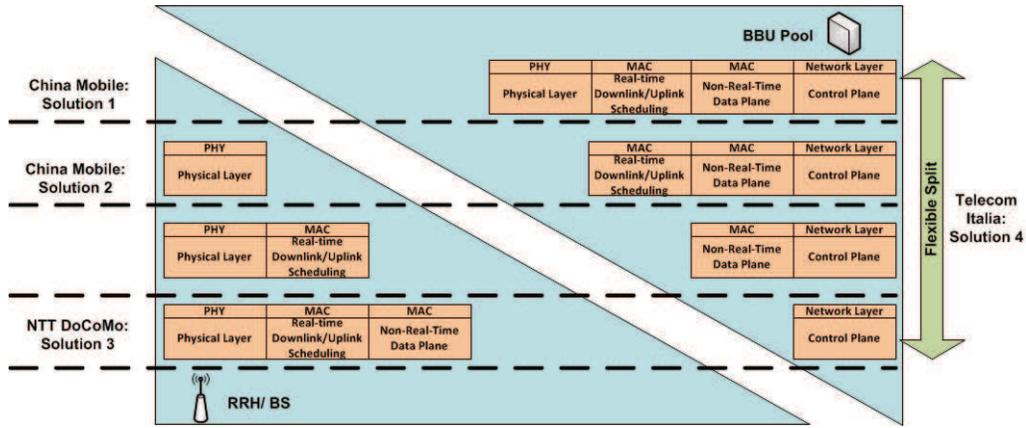}
\setlength{\belowcaptionskip}{-100pt} \vspace*{-10pt}
\caption{Functional split of C-RANs.}\label{split}
\end{figure*}

To take a tradeoff between implementing complexity and achieved LSCP gains, one of the key differences among those architectures advocated by the industry
lies in the degree of functional split between RRHs
and the BBU pool, as illustrated in Fig. \ref{split}.
For example, China Mobile Research Institute proposed fully-centralized
and partially-centralized architectures \cite{II:Mobile}.
The fully-centralized architecture, like Solution 1 in Fig. \ref{split}, shows a typical approach to deploy the C-RAN
, in which the functions in PHY, medium access control (MAC) and
network layers are all moved to the BBU pool. The main benefit of the fully-centralized architecture is that almost no digital processing devices are required at RRHs, potentially making them very small
and cheap, while the required data rate of fronthaul links for I/Q forwarding is comparatively high. The fully-centralized architecture can achieve the maximal LSCP gains, however, this kind of gains is achieved at the expense of a maximal heavy burden on fronthaul links because all I/Q signals of all UEs must be sent to the BBU pool timely. In \cite{s1}, the authors report that a required fronthaul capacity under the fully-centralized architecture is 10 Gbps for the LTE system with eight receive antennas with 20 MHz frequency bandwidth.

While in the partially-centralized architecture, like Solution 2 in Fig. \ref{split}, functions in PHY are integrated into RRHs, with other functions in upper layers incorporated at the BBU pool. Although the heavy burden on fronthaul links is alleviated in the partially-centralized architecture, LSCP can not be efficiently supported because functions in PHY remain in the individual RRH, and only the traditional distributed COMP gain can be achieved. Obviously, the required fronthaul capacity and signal processing capability in the BBU pool decrease significantly when the functional split is shifted to the MAC layer.

Furthermore, in~\cite{do}, NTT DoCoMo Company has introduced
an advanced C-RAN architecture to utilize carrier aggregation (CA) between marco and small cell carriers, in which most functions in PHY and MAC layers are
implemented in RRHs, and only upper-layer functions are moved to the BBU pool. The Solution 3 in Fig. \ref{split} shows that only the function of control plane remains in the BBU pool, which suggests only the CRRM gain can be obtained though the burden on fonthaul can be neglected \cite{front10}.

In contrast to the fixed functional split, the radio access network as a service (RANaaS) proposed by Telecom Italia Company allows a flexible split, like Solution 4 in Fig. \ref{split}, making it possible to choose an optimal operating point between the full centralization and local execution \cite{ita}, which is traded off with lower LSCP gains in terms of SE and the requirements of the high capacity of fronthaul. The advantage of a flexible split is to reap the benefits of both extremes: significant SE/EE gains for high requirements of fronthaul, and low SE/EE for capacity-free fronthaul.

\subsubsection{\textbf{Fronthaul Carrying Technologies}}

In addition, to fully boost the performance of C-RANs,
the fronthaul carrying technologies are extremely important, which can be passive optical network (PON), coax, cable, fiber, microwave, wireless communication, and even millimeter wave. When considering many RRHs connected to a common BBU pool in a general fully-centralized C-RAN architecture, the overall transport capacity for fronthaul will quickly extend into multiple 100 Gbps. The typical example of fronthaul protocol is CPRI, which has been defined with several options, such as option 1 (0.6144 Gbps), 2 (1.2288 Gbps), 3 (2.4576 Gbps), 6 (6.144 Gbps) and 7 (9.83 Gbps) with different carrying technologies~\cite{II:PONs}. Note that CPRI is not a standard, but an industry
agreement. In~\cite{II:Mobile}, four solutions for the fronthaul are considered: dark fiber, wavelength-division multiplexing (WDM) based PON, unified fixed radio access, and mobile radio access. Dark fiber is suitable when there are a lot of spare fiber resources, while the WDM based PON is preferred when fiber resources are limited, especially in the access ring.
In~\cite{II:RING}, a novel design algorithm of the access network with a ring topology
under the C-RAN architecture is proposed. Given the requirements of the fronthaul network segment, in~\cite{II:FRANCE}, a self-seeded
reflective semiconductor optical amplifier based technology for the WDM based PON is regarded as an attractive solution to C-RANs. An unified architecture for the fronthaul link, employing ring-based PON, is proposed in~\cite{II:WDM¨CPONs}. Considering the scarce fiber resources in some areas, ZTE Company introduced enhanced fiber connection and colored fiber connection as the bearer network solutions of C-RANs, in which only a few fibers are required to construct an independent C-RAN \cite{zte}. Specifically, the enhanced fiber connection provide a low bit rate transmission up to 10 Mbps, while the colored fiber connection can provide a high bit rate transmission up to 1 Gbps.

For the WDM based PON, there are many kinds of WDM techniques. The passive coarse wavelength division multiplexing (CWDM) is a cost efficient optical multiplexing technology, however, the fixed wavelength assignment and limited channel numbers in commercial CWDM solutions render it no much attraction. To tackle this problem, the dense wavelength division multiplexing (DWDM) is an attractive candidate, which offers dynamic management of the channel assignment. Accurately, fronthaul links need to meet strict latency and jitter requirements in order to synchronize the transmissions across massive RRHs. With the new NG-PON2 standard [4], DWDM is expected to be compliant with cost figures and operational needs for the capacity up to multiple 100 Gbps. NG-PON2 supports two kinds of DWDM solutions: 1) multiple unshared point-to-point connections via DWDM, and 2) multiple time division multiplexing /time division multiple access (TDMA) point-to-multipoint connections on a separate set of DWDM channels (e.g., time wavelength division multiplexing (TWDM))~\cite{II:NG-PON2}. It can be anticipated that TDMA based DWDM is a promising and preferred fronthaul carrier technology when the fronthaul capacity is unlimited and the cost of fronthaul is not a key factor for mobile operators.

\subsection{C-RAN Architectures Proposed by the Academia}

\begin{figure}[!h]
\centering \vspace*{0pt}
\includegraphics[scale=0.25]{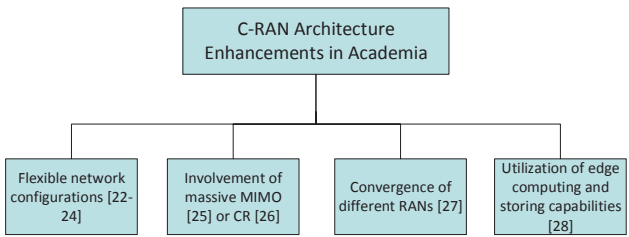}
\setlength{\belowcaptionskip}{-100pt} \vspace*{-10pt}
\caption{Different enhancements for C-RAN architectures in academia.}\label{aspects}
\end{figure}

The general C-RAN refers to the virtualization of base station functionalities by means of cloud computing, which can save the energy consumption, decrease the operation and maintenance cost, achieve the spatial processing gain, and adapt to the non-uniform traffic. However, the capacity-constrained fronthaul and the signal processing latency worsen the potential advantages of C-RANs. Meanwhile, there are some emerging problems to be solved from a system architecture viewpoint for making C-RANs rollout, including the convergence and interworking with the existing cellular networks, the advanced massive MIMO and CR techniques efficiently working in C-RANs, and the flexibility design for meeting the technical requirements of the fifth generation (5G) wireless systems. To tackle these challenges, much research has been carried out to enhance the general C-RAN architecture in the research community. As shown in Fig. \ref{aspects}, these academical works can be classified as the aspects of
flexible network configurations \cite{FluidNet,soft,das}, the involvement of promising techniques like massive MIMO \cite{large} and CR \cite{empowered}, the convergence of different RANs \cite{HCRAN}\cite{system}, as well as the utilization of edge computing and storing capabilities \cite{fog}.

\subsubsection{Flexible C-RANs}
The traditional C-RAN architecture is fully-centralized and fixed, which is not adaptive to the moveable traffic and the advanced software defined concept. As a result, it is urgent to improve the friable capability of C-RANs. In \cite{FluidNet}, the novel concept of re-configurable fronthaul is proposed to flexibly support
one-to-one and one-to-many logical mappings between BBUs and RRHs to
perform proper transmission strategies.
Specifically, for areas with static users and high traffic load,
one-to-one logical mapping between a BBU and a RRH will be configured to allow fractional frequency reuse
to meet the traffic demand. While for areas with mobile and low traffic, one-to-many mapping will be configured to allow
distributed antenna systems to deliver consistent coverage for mobile users and save the energy consumption of the
BBU pool. In \cite{soft}, the challenges of re-configurable fronthaul, also named as software defined fronthaul (SDF), is further discussed, in terms of latency, communication protocol, and so on.
In addition, the authors in \cite{das} introduced a software defined architecture for C-RANs,
which can be implemented on general purpose processors. With this architecture, the C-RAN can be flexibly configured to operate as
different networks.

\subsubsection{C-RANs with Massive MIMO and CR}

To further improve SE and EE with advanced techniques, some researchers have proposed to use the massive MIMO and CR techniques in C-RANs.
In \cite{large}, the massive MIMO technique is involved.
By exploiting the proper transmit precoding or receive combining, hundreds of terminals can be simultaneously served by one RRH, substantially improving
the network capacity. However, under a fully centralized C-RAN,
when the number of antennas at RRHs becomes large, the fronthaul has to face a huge volume of data.
To alleviate the fronthaul constraint while maintaining the collaborative processing gain,
the functionalities in PHY can be divided into a centralized part and a distributed part.
Particularly, in the downlink, precoding vectors and data symbols for
the scheduled users are handled by the BBU pool, while
RRHs precode the user symbols and transmit radio signals.
The authors in \cite{empowered} studied the application of CR to the C-RAN,
where RRHs are capable of sensing the spectrum and send the related information to
the BBU pool. The BBU pool then assigns the proper frequency for cell sites based on the received information,
which can effectively mitigate the interference thus boost the system capacity.

\subsubsection{Other Enhanced Architectures}

The traditional C-RAN is proposed to improve SE and EE, in which the control signal design is not optimized, and the interworking with the existing cellular networks is not considered as well. Focusing on taking full advantages of both heterogenous networks and C-RANs, the authors in \cite{HCRAN}\cite{system} presented an H-CRAN as the advanced
wireless access network paradigm, in which a new
communication entity named as Node C (Node with cloud computing) is defined.
When this entity is used to converge macro BSs (MBSs), micro BSs, and other heterogeneous nodes,
it can be seen as a convergence gateway. When it is used to manage RRHs, it acts as a BBU pool.
Meanwhile, four enabling techniques, including the advanced spatial signal processing,
cooperative radio resource management, network function virtualization, and self-organizing network, are proposed
to fulfill the potential of H-CRANs.
Although several solutions have been proposed to alleviate the burden over the fronthaul,
the fronthaul in H-CRANs still suffers from the explosive traffic.
To save fronthaul resource consumption and reduce delay, an F-RAN architecture
is designed in \cite{fog}, in which a large number of collaboration radio signal processing
and cooperative radio resource management functions can be executed in
fog computing based access points or by ``smart'' UEs, and some packets
can be directly delivered by the edge devices through their limited cache.

\subsection{C-RANs towards 5G}

It is envisioned that 5G will bring a 1000x increase in terms of area capacity compared with 4G,
achieve a peak rate in the range of tens of Gbps, support a roundtrip latency of about 1 ms
as well as connections for a trillion of devices, and guarantee ultra reliability \cite{ad,no}.
In \cite{s1}, the field trials conducted by China Mobile have verified the throughput gain
brought by C-RANs based on an uplink LTE model, reaching up to near 300\%. Through dense RRHs in C-RANs, massive connections are efficiently supported, and it is not hard to provide good service for trillion of devices if the density of RRHs is sufficiently high.
Although a big gap is still observed compared to 5G requirements, the result has shown the potential advantages of C-RANs.
Meanwhile, different advanced techniques
can be involved in C-RANs to further improve the spectrum efficiency, including CR, massive MIMO, and full duplex radio.
Moreover, mmWave spectrum can be used to provide much available bandwidth.
As a result, it can be anticipated that C-RANs will be qualified for meeting the high SE and EE in 5G.

The traditional C-RAN architectures should be enhanced to meet low latency and high reliability requirements.
For example, the fully centralized C-RAN architecture proposed in \cite{II:Mobile} puts all functions of the air interface at the
BBU pool.
Hence, the centralized C-RAN architecture is liable to cause long latency, and decrease the reliability when the fronthaul is constrained.
Thus, decentralization is an alternative. Recently, in academia,
with the integration of remote clouds and distributed cloudlets, fiber-wireless access networks can
improve cloud accessibility with low latency and high reliability \cite{fiwi}. Meanwhile in industry,
Nokia has proposed to change the role of mobile BSs by endowing them with
some IT-based capabilities like localized processing and content storage to implement distributed radio access networks (D-RANs) \cite{in}.

\section{Fronthaul Compression}

In recent years, next-generation passive optical
network (NG-PON) has been considered one of the most promising optical access technologies, which is envisioned to support a high data rate (over 10 Gbps). To save the building cost, the popular and industrial optical access technologies are Ethernet PON (EPON) and Gigabit EPON (GEPON), which take advantage of inexpensive and ubiquitous Ethernet equipment, and offer the transmission speed up to 1 Gbps and 10 Gbps, respectively \cite{s5}. The intuition of the performances of different fronthaul technologies has been illustrated in Table \ref{front}.

\begin{table}[h]
\caption{The Performances of Different Fronthaul Technologies}\label{front}
\scriptsize 
\centering
\begin{tabular}{|p{0.7 in}|p{0.7 in}|p{0.7 in}|p{0.7 in}|}
 \hline
 \textbf{Fronthaul Technology}& \textbf{One-way Latency} &\textbf{Per-hop Latency}& \textbf{Throughput} \\ \hline
 NG-PON & 2.5 $\mu s$ &5 $\mu s/km$ & Over 10 $Gbps$ \\ \hline
 GE-PON & 10-30 $ms$ & 1 $ms$ & 10 $Mbps$-10 $Gbps$ \\ \hline
 EPON & 5-10 $ms$ &1 $ms$ &10 $Mbps$-1 $Gbps$ \\ \hline
 Digital Subscriber Line Access & 5-35 $ms$ &5-35 $ms$ &10 $Mbps$-100 $Mbps$ \\ \hline
 Cable & 25-35 $ms$ &25-35 $ms$ &10 $Mbps$-100 $Mbps$ \\ \hline
 Wireless Communication (200 MHz - 6 GHz) & 5-10 $ms$ &5 $ms$ &50 $Mbps$-1 $Gbps$ \\ \hline
 Microwave & $< $ 1 $ms$ &200 $\mu s$ & 100 $Mbps$-1 $Gbps$ \\ \hline
 Millimeter wave radio & $< $ 1 $ms$ &200 $\mu s$ & 500 $Mbps$-2 $Gbps$ \\ \hline
\end{tabular}
\end{table}

In C-RANs, the fronthaul links carry
information about the baseband signals in the form of quantized I/Q samples, and the large
bit rates produced by the
quantized I/Q signals contradicts the limited capacity of the practical fronthaul solutions like the dark fiber solution \cite{s1},
which has a significant impact on the LSCP gains \cite{s4}. In the uplink, for instance, RRHs need to sample, quantize, and then forward the received I/Q signals to the BBU pool. With densely deployed RRHs, the fronthaul traffic generated from a single UE with several MHz bandwidth could be easily scaled up to multiple Gbps. In practice, a commercial fiber link with tens of Gbps capacity could thus be easily overwhelmed even under moderate mobile traffic. Therefore, to tackle this problem, one approach is to utilize the partial centralization structure though substantial signal processing capabilities are required on RRHs. The other alternative is to adopt advanced techniques to optimize the performance under a fully centralized structure with constrained fronthaul. As it simplifies the functions and capabilities of RRHs, the latter solution is the focus of current research community, and one of the corresponding key techniques is the fronthaul compression \cite{s4}. The aim of fronthaul compression is to compress the I/Q rate across multiple RRHs for all served UEs adaptive to the available fronthaul capacity, which is critical to alleviating the impact of fronthaul capacity constraints.

Generally speaking, current fronthaul compression techniques can be classified into
the quantization based compression, compressive sensing (CS) based compression, and spatial filtering.
The first one mainly includes point-to-point compression and distributed source coding in
uplink C-RANs as well as point-to-point compression and joint compression in downlink
C-RANs. While the latter two are mainly used in uplink C-RANs. For the quantization based compression,
the compression is achieved by quantizing original signals, and the variance of the quantization noise, often assumed to be
zero-mean complex Gaussian, should be optimized. For the CS based compression and spatial filtering, the compression is performed by multiplying received signals with a local compression
matrix at each RRH, which can reduce the dimension of the signals to be transmitted to the BBU pool. Note that the distributed source coding and joint compression are often called by network-aware compression technique. The core idea of distributed source coding is to enable the BBU pool to leverage the correlation among the signals
received by neighboring RRHs, while CS is to leverage the sparsity of the UEs' signals. Hence, such techniques are
both available for uplink and are not directly applied in downlink.

\subsection{Uplink Compression}

\begin{figure}[tp]
\centering
\includegraphics[scale=0.45]{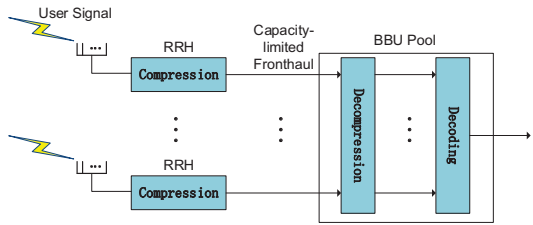}
\caption{The uplink compression model in C-RANs.}\label{u}
\end{figure}

In the uplink, each RRH compresses its
received baseband signal and forwards the compressed
data to the central BBU pool as a soft relay
through a limited-capacity fronthaul link as shown in Fig. \ref{u}. To achieve uplink fronthaul compression, there are four main approaches: point-to-point compression \cite{III-2}, distributed
source coding, CS based compression, and spatial filtering.
When the low implementation complexity is preferred, the point-to-point compression and spatial filtering are recommended. When
signals received at different RRHs are closely correlated, the distributed source coding should be used.
Considering the sparsity of the uplink C-RAN signals, CS based compression is an another alternative.
At the BBU pool,
the joint decoding operation based on all quantization values from all RRHs is executed.
The state of the art of uplink fronthaul compression is briefly summarized in Table \ref{fro1}.

In \cite{III-4}, it is observed that the imperfect knowledge of the joint statistics
of the signals received at RRHs can lead to a performance degradation of distributed
source coding. To deal with this problem, a robust distributed compression scheme
is proposed. Moreover, the network EE is taken into account by
addressing RRH selection, in which an optimization
problem aiming at the sum rate maximization is formulated by introducing a sparsity-inducing term to
jointly consider the compression and RRH selection.
Simulation results show the effectiveness of the robust scheme on compensating the
performance loss due to the imperfect statistical information. Still tackling
the issue of robustness in distributed compression, the authors of \cite{II:Simeone2} investigated the layered
transmission and compression strategies.
Under competitive robustness and fronthaul capacity constraints,
the compression is formulated as the minimization of the transmit
power. It is shown that the proposed layered
strategies greatly outperform single-layer
strategies, and it is the layered transmission not the layered
compression that contributes the main performance gains.

Authors in \cite{III-6}
formulated the problem of optimizing the quantization noise
levels to maximize weighted sum rate under a sum fronthaul
capacity constraint by involving the distributed Wyner-Ziv coding and single-user compression.
Different from \cite{III-4}, in which the
quantization noise level optimization problem is solved on a
per-RRH basis, \cite{III-6} jointly optimized
the quantization noise levels of all RRHs. It is shown that setting the
quantization noise levels to be proportional to the background
noise levels is near optimal for sum-rate maximization when the
signal-to-quantization-noise-ratio (SQNR) is sufficiently high.
Furthermore,
it is concluded that the gap between the sum rate achieved by Wyner-Ziv coding based scheme
and a cut-set like sum capacity
upper bound is bounded by a constant when quantization noise levels are set to be proportional to the background
noise levels. With single-user compression, a similar result exists
under a diagonally dominant channel condition.

Although the fronthaul loading can be effectively reduced by proposals in \cite{II:Simeone2,III-6},
the uplink signal sparsity structure is ignored. In \cite{III-7},
by exploiting the signal sparsity in uplink C-RANs
, the distributed CS and recovery
techniques are used to compress the fronthaul loading distributively.
The performance of the proposed end-to-end signal recovery algorithm
is analyzed, and it is shown that the restricted isometry property
can still be satisfied by the aggregate measurement
matrix in C-RANs containing both distributed fronthaul compression
and multi-access fading.
While in \cite{III-order}, to realize a better design of the cloud decoder,
joint decompression and decoding are performed at the BBU pool.
The sum-rate maximization problem is a difference of convex problem.
By numerical results, the advantage
of the proposed algorithm is demonstrated, compared to the conventional
approach based on separate decompression and decoding as in \cite{III-4}.

It is worth noting that in the presence of time-varying channels, the overhead of communicating
the CSIs from RRHs to the BBU pool can becomes sufficiently large. Thus it is
essential to study the transfer of CSIs and data from RRHs to the BBU pool. In \cite{III-9},
the research on joint signal and CSI compression is conducted. Motivated by results in \cite{III-10}
related to the separation of estimation and compression, an estimate-compress-forward (ECF) approach is deployed, in which
CSIs are first estimated at each RRH, and then the estimated CSIs are compressed for the transmission to the BBU pool.
Based on ECF approach, various strategies for the
separate or joint compression of the estimated CSIs and the received data signal are proposed and analyzed.
When there are multiple RRHs, the proposed strategies can be combined with distributed
source coding to leverage the received signal correlation
across RRHs. It is observed that the ECF approach outperforms the compress-forward-estimate
approach as used in \cite{III-11}.

In \cite{front9}, the authors considered the uplink of a C-RAN with multi-antenna RRHs each connected to
the BBU pool through a finite-capacity fronthaul link, and proposed a novel spatial-compression-and-forward (SCF) scheme.
Unlink literatures \cite{III-4,III-6} adopting complex quantization schemes, a simple linear filter is used
to compress the correlated signals received by all the antennas at each RRH.
An optimization problem is formulated, aiming at maximizing the
minimum SINR of all the users by jointly optimizing users' power allocation, RRHs'
spatial filter design and quantization bits allocation, and BBU' s receive beamforming.

Whereas in \cite{front5}, the quantization strategy optimization is handled for the
positioning in C-RANs, and its goal
is to minimize the worst-case localization error of a single-antenna radio transmitter under fronthaul capacity constraints.
The Charnes-Cooper transformation and difference-of-convex programming are utilized to design corresponding
optimization algorithms, and several conclusions are got by numerical simulations.
For example, it is observed that a larger
fronthaul capacity brings an improved localization, and
the localization has lower fronthaul rate requirements compared with data communication.

Most of the aforementioned works only considered the scenario where RRHs are directly connected to the
BBU pool. Instead, in \cite{III-12}, a general multi-hop fronthaul scenario is studied, in which
each RRH may communicate with the BBU pool through a set of intermediate RRHs. In this scenario, the
multiplex-and-forward (MF) strategy may suffer from a significant performance loss when RRHs are densely deployed.
To solve this problem, a decompress-process-and-recompress (DPR) strategy is proposed,
in which each RRH decompresses the received streams and performs linear in-network processing of the decompressed signals.
For both MF and DPR strategies, the optimal design is handled to maximize the sum-rate under fronthaul capacity constraints.
By comparing the performance of MF and DPR strategies, the advantage of in-network processing is highlighted.

\begin{table*}[!htp]
\caption{SUMMARY OF LITERATURES RELATED TO UPLINK FRONTHAUL COMPRESSION}\label{fro1}
\scriptsize 
\centering
\begin{tabular}{|p{0.5 in}|p{0.6 in}|p{1.93 in}|p{0.6 in}|p{2 in}|} 
 \hline
 \textbf{Literature}& \textbf{Compression Method} &\textbf{System model}& \textbf{Optimization objective}& \textbf{Main contribution} \\ \hline
 \cite{III-4} & Distributed source
coding&Multiple multi-antenna RRHs and multi-antenna users scenario with per-RRH fronthaul capacity constraint&Maximize the sum rate & Propose a
robust distributed compression scheme that can compensate the
performance loss due to the imperfect statistical information \\ \hline
 \cite{III-6} & Distributed source coding and point-to-point compression&Multiple single-antenna RRHs and single-antenna users scenario with a sum fronthaul capacity constraint& Maximize the weighted sum rate&
Find near optimality of setting the
quantization noise levels to be proportional to the background
noise levels for sum-rate maximization under high SQNR \\ \hline
 \cite{III-7} & CS based compression &Multiple single-antenna RRHs and single-antenna users scenario without fronthaul capacity constraint & Maximize the probability of correct active user detection & Exploit the signal sparsity in the uplink C-RAN and analyze the tradeoff relationship between the
C-RAN performance and the fronthaul loading \\ \hline
 \cite{III-12} & Point-to-point compression&Multiple multi-antenna RRHs and multi-antenna users scenario with
a multihop fronthaul topology & Maximize the sum rate & Propose a DPR strategy and highlight
the advantage of in-network processing \\ \hline
\cite{front9} & Spatial filtering & Multiple multi-antenna RRHs and single-antenna users scenario with
per-RRH fronthaul capacity constraint & Maximize the
minimum SINR of all users & Show that a multi-antenna C-RAN generally significantly outperforms
both massive MIMO and a single-antenna C-RAN under practical fronthaul rate constraints \\ \hline
\cite{front5} & Point-to-point compression& Multiple single-antenna RRHs and a single-antenna user scenario with
per-RRH fronthaul capacity constraint & Minimize the worst-case localization error of the target &
Propose optimization algorithms based on Charnes-Cooper transformation and difference-of-convex programming\\ \hline
\end{tabular}
\end{table*}

\subsection{Downlink Compression}
\begin{figure}[tp]
\centering
\includegraphics[scale=0.4]{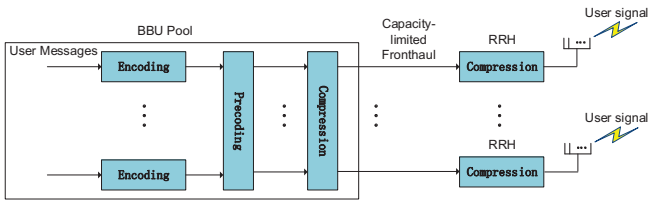}
\caption{The downlink compression model in C-RANs.}\label{III-downlink}
\end{figure}

Fig. \ref{III-downlink} shows an example of the downlink compression in C-RANs. First, the BBU pool performs
channel coding and pre-coding for users' message, and before transmission on
the fronthaul, the pre-coded signals are compressed jointly or separately. Then RRHs decompress the received signals and
transmit them to the desired users.
The state of the art of downlink fronthaul compression is briefly surveyed,
and the main literatures are summarized in Table \ref{fro2}.

\subsubsection{Compression under Static Channels}

In \cite{II:Simeone3}, multivariate compression is leveraged to suppress
the additive quantization noises at UEs.
The joint pre-coding and compression design problem to maximize the weighted sum-rate is formulated under power and fronthaul capacity constraints, and
a stationary point is reached by a proposed iterative algorithm.
Moreover, the practical implementation of the multivariate compression and the robust design of the pre-coding and compression are tackled as well.
Numerical results have demonstrated that
the proposed joint precoding and compression
strategy outperforms conventional approaches based on
the separate design of precoding and compression.
However, the work \cite{II:Simeone3} has its own limitations that the optimization of the pre-coding
and the joint compression is only within a given cluster of RRHs
without considering the effect of inter-cluster interference. When applied
to scenarios of multiple mutually interfering clusters, inter-cluster interference could degrade the performance of its approach.
Based on this observation, authors in \cite{III-14} investigated
the joint design of pre-coding
and fronthaul compression in a multi-cluster case,
in which the problem of maximizing
the weighted sum-rate across all clusters
is formulated, and an iterative algorithm is proposed to find the solution to the problem.
The simulation results show that a significant gain can be reaped by the proposed scheme.

Both in \cite{II:Simeone3} and \cite{III-14}, the joint pre-coding is executed at the BBU pool,
which is classified into the pure compression strategy. The pure compression means that the precoded analog signals are compressed and forwarded to the corresponding RRHs in downlink, which is akin to the compress-and-forward relaying strategy. To implement the joint pre-coding,
a pure message-sharing strategy \cite{III-15} can be used, in which the BBU pool
simply shares each user's message with multiple RRHs over the
fronthaul links. The pure message-sharing strategy can be thought of as analogous to the decode-and-forward relaying strategy. Through combining advantages of both pure compression and pure message-sharing, a hybrid strategy adaptive to the available fronthaul capacity is proposed in \cite{III-15}, whose aim is to allocate some fronthaul capacity to send direct messages for some UEs (for whom RRHs are better off receiving messages directly, instead of their contributions in the compressed precoded signals) by the pure message-sharing strategy, and the remaining fronthaul capacity is used to carry the compressed signal that combines the contributions from the rest of UEs by the pure compression strategy. The analysis and simulation results show that the hybrid strategy is more beneficial to the overall system performance in a practical C-RAN with finite fronthaul capacity, compared with any separated compression strategy.

\subsubsection{Compression under Dynamic Channels}

Different from literatures \cite{II:Simeone3,III-14,III-15 } assuming
static channels and ideal CSIs at the BBU pool, the authors in
\cite{front6} investigated the joint fronthaul compression and precoding design
with a block-ergodic fading channel model in the downlink C-RAN with a cluster of
RRHs with multiple antennas serving multi-antenna users.
The analysis for two types of BBU-RRH functional splits at PHY are conducted,
which corresponds to the compression-after-precoding (CAP) and compression-before-precoding (CBP) strategies, respectively.
For the first strategy, all the baseband processing is performed at the BBU pool, while for the second strategy,
RRHs perform channel encoding and precoding instead of the BBU pool which only separately forwards the
user message and the compressed precoding matrices to each RRH.
The ergodic capacity optimization for both strategies is tackled by
the proposed stochastic successive
upper-bound minimization approach. The numerical results show
that the optimal BBU-RRH functional split
depends on the interplay between the enhanced
interference management abilities of CAP and the lower fronthaul requirements of
CBP.

\begin{table*}[!htp]
\caption{SUMMARY OF LITERATURES RELATED TO DOWNLINK FRONTHAUL COMPRESSION}\label{fro2}
\scriptsize 
\centering
\begin{tabular}{|p{0.5 in}|p{0.6 in}|p{2.13 in}|p{0.6 in}|p{1.8 in}|}
 \hline
 \textbf{Literature}& \textbf{Compression Method} &\textbf{System model}& \textbf{Optimization objective}& \textbf{Main contribution} \\ \hline
 \cite{II:Simeone3} & Joint
compression &A given cluster composed of multiple multi-antenna RRHs and multi-antenna users with per-RRH fronthaul capacity constraint& Maximize the weighted sum rate & Suppress the additive quantization noises at users by leveraging multivariate compression\\ \hline
 \cite{III-14} & Joint
compression &Multiple mutually interfering clusters each of which consists of multiple multi-antenna RRHs and multi-antenna users with per-RRH fronthaul capacity constraint& Maximize the weighted sum rate&
 Joint design of pre-coding
and fronthaul compression in a multi-cluster case\\ \hline
 \cite{III-15} & Point-to-point compression& Multiple single-antenna RRHs and single-antenna users scenario with per-RRH fronthaul capacity constraint & Maximize the weighted sum rate & Propose a hybrid scheme of pure compression and pure message-sharing, achieving significant performance gains \\ \hline
 \cite{front6} & Point-to-point compression& Multiple multi-antenna RRHs and multi-antenna users scenario with per-RRH fronthaul capacity constraint & Maximize the ergodic capacity & Joint fronthaul compression and precoding design
with a block-ergodic fading channel model, which yields insights into the optimal BBU-RRH functional split for C-RANs \\ \hline
\end{tabular}
\end{table*}

\subsection{\textcolor{red}{Lessons Learned}}

Due to the constrained fronthaul and decreasing the overhead of transmitting I/Q signals, the fronthaul compression is key for both uplink and downlink in C-RANs. Point-to-point compression can reduce the bit rate of fronthaul through compression as applied separately on each fronthaul link, while the network-aware compression can provide a significant compression gains, which towards the network information-theoretic optimal performance. In uplink, the network-aware compression techniques include the distributed source coding and spatial filtering. While in downlink, since the point-to-point compression is suboptimal from a network information-theoretic viewpoint, the joint multivariate compression is preferred. Comparing with downlink, the uplink compression is more challenging because I/Q signals from UEs are sparsely and massively distributed. Generally, the network-aware compression technique outperforms the separate design of precoding and compression. Compared with the separated strategy \cite{III-15}, the hybrid of pure compression and message-sharing is more beneficial to the overall system performance in a practical C-RAN with finite fronthaul capacity. In the future, the network-aware compression and the sparse signal processing should be jointly optimized to alleviate the fronthaul constraint. Meanwhile, the network-aware compression requires the full and ideal CSIs of both the access links between RRHs and UEs and the wireless fronthaul links, which are interesting open problems in uplink.

\section{Large-scale Collaborative Processing}

\begin{figure}[!h]
\centering \vspace*{0pt}
\includegraphics[scale=0.4]{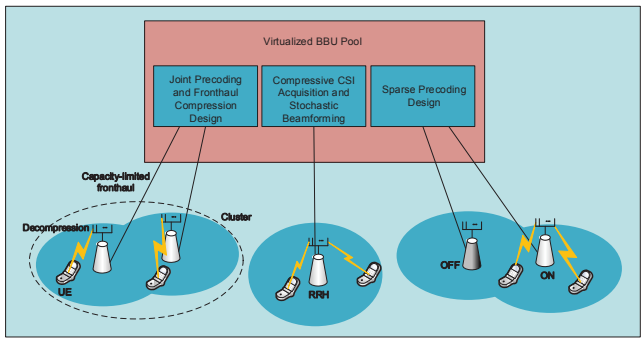}
\setlength{\belowcaptionskip}{-100pt} \vspace*{-10pt}
\caption{Different precoding designs in C-RANs.}\label{precoding}
\end{figure}

Without advanced interference coordination techniques in C-RANs,
interference should be severe due to the dense deployment of RRHs, which
significantly decreases SE and EE of C-RANs.
To mitigate the interference and thus improve SE and EE, the centralized LSCP techniques
should be executed in the BBU pool.
Currently, several approaches have been proposed
to solve this challenge as shown in Fig. \ref{precoding},
including the joint precoding and fronthaul compression design to alleviate the fronthaul capacity constraint,
the compressive CSIs acquisition and stochastic beamforming to deal with the difficulty in obtaining perfect CSIs, and
the sparse precoding design to address RRH selection. In this section,
the related papers are summarized in Table \ref{la}.

\begin{table*}[!htp]
\caption{SUMMARY OF LITERATURES RELATED TO LARGE-SCALE COLLABORATIVE PROCESSING}\label{la}
\scriptsize 
\centering
\begin{tabular}{|p{0.6 in}|p{0.6 in}|p{1.6 in}|p{1 in}|p{1.8 in}|}
 \hline
 \textbf{Literature}& \textbf{Classification} &\textbf{Problem Description}& \textbf{Challenge} & \textbf{Solution} \\ \hline
 \cite{IV:MCP} & Precoding with perfect CSIs & Joint RRH selection and linear precoding design for the downlink transmission in networks with multi-cell processing &High complexity for RRH selection & RRH selection with the guidance of the sparsity pattern of the precoding matrix \\ \hline
 \cite{II:Simeone3} & Precoding with perfect CSIs & Joint precoding and fronthaul compression design for downlink C-RANs &
 The control of the effect of additive quantization noises at the MSs & Multivariate compression \\ \hline
 \cite{III-14} & Precoding with perfect CSIs & Inter-cluster design of precoding and fronthaul compression & The existence of mutually interfering clusters
 & Inter-cluster optimization of
precoding and joint compression by a proposed iterative algorithm\\ \hline
 \cite{IV:SCB1} & Precoding with imperfect CSIs & Beamforming design for a downlink Cloud-RAN & Huge overhead to obtain full CSIs & Compressive CSIs acquisition and stochastic beamforming\\ \hline
 \cite{ro} & Precoding with imperfect CSIs & MMSE precoding algorithm for a downlink Cloud-RAN & Channel estimation errors and low-complexity algorithm design& Bayesian philosophy for reducing the effects of channel estimation errors and alternating direction method of multipliers for decreasing computational complexity \\ \hline
 \cite{netpo} & Precoding with imperfect CSIs & Beamforming for multicast green Cloud-RAN & An infinite number of the non-convex
 quadratic QoS constraints & The adoption of convex optimization technique based on PhaseLift, semidefinite relaxation and S-lemma \\ \hline
 \cite{IV:yu2013} & Sparse precoding & Beamforming for limited-fronthaul network MIMO system &
The non-convexity of the problem of finding the optimal tradeoff between the total transmit power and the sum fronthaul
capacity & Iteratively relaxing the $\mathop l\nolimits_0 $-norm as a weighted $\mathop l\nolimits_1 $-norm \\ \hline
\cite{IV:yu2014} & Sparse precoding & Joint beamforming and clustering design for a downlink network MIMO system & Per RRH fronthaul constraints with
$\mathop l\nolimits_0 $-norm & Iteratively approximating the per RRH fronthaul constraints by a reweighted $\mathop l\nolimits_1 $-norm technique \\ \hline
\cite{IV:pre10} & Sparse precoding & Joint RRH selection and power minimization beamforming for a Cloud-RAN & High complexity for solving the formulated problem & Group sparse beamforming method based on the group-sparsity of beamformers induced by the weighted ${{\mathop l\nolimits_1 } \mathord{\left/
 {\vphantom {{\mathop l\nolimits_1 } {\mathop l\nolimits_2 }}} \right.
 \kern-\nulldelimiterspace} {\mathop l\nolimits_2 }}$-norm minimization \\ \hline
\end{tabular}
\end{table*}

\subsection{Precoding with Perfect CSIs}

Several literatures have been
published for the precoding design recently,
most of which are presented to target the mitigation of interference and the improvements of SE and EE performances. Dirty paper
coding (DPC)~\cite{IV:pre1}~\cite{IV:MHM} achieves capacity region, however, the complexity is too high
especially for large-scale C-RANs. Therefore, there is
few published works considering DPC for the large-scale C-RANs. On the other hand,
the linear precoding is regarded as a low complex alternative~\cite{IV:pre3,IV:pre4,IV:pre5,IV:pre6}.
The linear precoding techniques mainly utilize CSIs in a simple and efficient way to manage interference
across users and among the data streams for the same user.

\subsubsection{Precoding Design with Limited Fronthaul Capacity}

Considering the fronthaul capacity is limited in practice and the load increases with the number of
cooperating RRHs, a joint RRH selection and linear precoding
design algorithm is investigated in~\cite{IV:MCP},
in which the sparsity pattern of the precoding matrix gives some information on RRH selection.
For previous RRH selection schemes, RRHs with shorter distances from mobile stations or with better channel conditions
are chose. Differently,
the algorithm proposed in~\cite{IV:MCP} selects the cooperating RRHs based on the regularized convex optimization.
Numerical results
suggest that the proposed algorithm can achieve a significant
performance gain with a low computational
complexity.

To further overcome the indispensable limitation of fronthaul capacity,
the authors in~\cite{II:Simeone3} studied the joint design of precoding
and compression strategies for the downlink C-RAN with limited fronthaul
capacity.
DPC and linear precoding are compared, and the numerical results
show that DPC has better performance only in the regime of intermediate transmit power due to
the limited-capacity fronthaul links, which suggests that the compression has bigger impact
on the system performance rather than the precoding methods. However, the paper mainly focuses
on the optimization of precoding coefficient design within a single cluster of RRHs, and the
corresponding performance gain is marginal if considering interference from other clusters. To
this end, the authors of \cite{III-14} considered a weighted sum rate
maximization problem subject to fronthaul capacity and per RRH power constraint with the aim of
optimizing the precoding and quantization covariance matrices across all clusters.
An iterative algorithm based on the majorization minimization approach
is utilized to achieve a stationary point, in which the inter-cluster design, inter-cluster time-division multiple access and intra-cluster design algorithms are proposed to enable
the achievable rates to approach the near-optimal configuration. It
is learned the proposed
precoding based on multivariate compression and inter-cluster design significantly outperforms
the intra-cluster approach in~\cite{II:Simeone3}.

\subsubsection{Precoding Design for EE Maximization}

Another aspect is to maximize the EE performance.
Due to the huge power consumption caused by a large number of RRHs and
fronthaul links, the power consumption has a significant
impact on EE~\cite{IV:pre10}.
Considering the fact that EE is one of the major objectives in the future
cellular networks, allowing some RRHs to enter sleep mode
is beneficial when the traffic load is not huge in the
corresponding area ~\cite{IV:pre11}.
For the power consumption of the fronthaul links, it is
determined by the set of active RRHs, and the transmit power consumption of the
active RRHs can be minimized through the coordinated beamforming. As a result, a joint RRH selection
and power minimization problem is
built in~\cite{IV:pre10}, in which the partial precoding design is proposed by applying optimization
with group sparsity induced norm. The sparsity patterns
in signal processing have been exploited for a high efficient
system design~\cite{IV:pre12}. The $l1$-norm regularization has been successfully applied in
compressed sensing~\cite{IV:pre13,IV:pre14}.
On the other hand, the mixed $l1/l2$ -norm~\cite{IV:pre15} and
$l1/l\infty$-norm~\cite{IV:pre16} are most commonly used to induce the group sparsity because
of their analytical and computational convenience.

\subsection{Precoding with Imperfect CSIs}

\subsubsection{Precoding with Statistical CSIs}

Considering the BBU pool can typically
support hundreds of RRHs, it is apparent that the acquisition of full CSIs is critical to the
optimal design of precoding in C-RANs. Note that many works commonly assumed the BBU obtains the full CSIs through pilot training, however, obtaining full CSIs
is challenging because a lot of parameters are involved, leading to
significant estimation errors, quantization errors, and long feedback delays~\cite{IV:pre17}. Fortunately, partial CSIs can be more accurately obtained
through long term channel feedback. Precoding based on the statistical
CSIs has been well studied for single-user multiple input multiple output (SU-MIMO)~\cite{IV:pre18,IV:pre19} and then extended
to multiuser MIMO (MU-MIMO)~\cite{IV:pre20,IV:pre21}. A scheme firstly proposed
in~\cite{IV:pre22} considers
a transmission scheme that is proved to achieve a larger degree of freedom
than one with completely outdated CSIs. This scheme alongside
precoding is used to improve the achievable rate in~\cite{IV:pre23}. On this account,
the statistical CSIs in conjunction with this transmission scheme is
utilized in~\cite{IV:pre17} to develop both
transmission and precoding schemes that
provide further system performance improvement.

\subsubsection{Stochastic Coordinated Beamforming}

To overcome the curse of dimensionality for C-RANs and provide an CSIs overhead reduction, the authors in~\cite{IV:SCB1} proposed a novel CSIs acquisition scheme called compressive CSIs acquisition, which can obtain
the instantaneous CSIs of a subset of channel links and
the statistical CSIs of the others. Through this scheme, more overhead can be saved due to the reduced channel compression dimension in large-scale C-RANs.
With these mixed CSIs, a coordinated beamforming
is conducted to minimize the total transmit power while satisfying the QoS requirement for all UEs. The authors proposed a stochastic coordinated beamforming (SCB) framework based on the chance-constrained programming by exploiting the statistical CSIs to guarantee the QoS requirements. It is revealed that
the proposed SCB scheme can significantly reduce the overhead while providing performance close to that with full CSIs.
However, it is noted that the required number of samples
increases rapidly with the size of networks.
Given the fact that ~\cite{IV:SCB1} only provides some feasible solutions without any optimality guarantee, authors in~\cite{IV:SCB2} proposed a novel stochastic
difference-of-convex programming algorithm to solve this SCB problem, and it can reach the globally optimal solution if the problem is convex and reach a locally
optimal solution if the problem is non-convex. It is learned that the CSIs acquisition
overhead can be reduced by about 40\%.

\subsubsection{Robust Precoding Design}

In \cite{ro}, a robust low-complexity minimum mean square error precoding design problem is investigated,
aiming at minimizing the total average mean square error with per-antenna power constraint.
Here, imperfect CSIs is concerned due to the Gaussian distributed channel estimation errors.
The alternating direction method of multipliers (ADMM) is employed to decouple the optimization objective and the constraints,
greatly decreasing the complexity of the proposed robust algorithm, especially for the large-scale C-RANs.
Simulation results show that the proposed ADMM-based robust design can improve the system performance.

\subsubsection{Precoding Design for Energy Minimization}

The authors in \cite{netpo} focused on the energy minimization problem for the multicast C-RAN under imperfect CSIs,
which includes an infinite number of non-convex quadratic QoS constraints.
A computationally efficient and robust algorithm with three stages is proposed.
In the first stage, a novel quadratic variational formulation of
the weighted mixed ${{\mathop l\nolimits_1 } \mathord{\left/
 {\vphantom {{\mathop l\nolimits_1 } {\mathop l\nolimits_2 }}} \right.
 \kern-\nulldelimiterspace} {\mathop l\nolimits_2 }}$-norm is used to induce the
group-sparsity structure for the robust multicast beamformer for the guidance of RRH selection.
The perturbed alternating
optimization algorithm is then considered to solve the resultant non-convex group-sparsity
inducing optimization problem.
In the second stage, a PhaseLift method is introduced to handle
the feasibility problems in the RRH selection procedure, and
the semidefinite relaxation technique is adopted to design the beamformer in the final stage.

\subsection{Sparse Precoding}
In C-RANs, the power consumption of fronthaul
has a significant impact on EE. Hence, allowing the fronthaul links and the corresponding RRHs to support the sleep mode are essential
to reduce the power consumption for C-RANs. It would be feasible to switch off some
RRHs while still maintaining the QoS
requirements. Group sparsity is required rather than the individual sparsity as used in
compressed sensing~\cite{IV:pre13}~\cite{IV:pre14}, similar
to~\cite{IV:pre15}~\cite{IV:pre16}. In~\cite{IV:pre10}, a mixed $l_1/l_p$-norm is adopted
to induce sparsity in large-scale C-RANs, however,
the approach in~\cite{IV:pre15}~\cite{IV:pre16} simply considered
un-weighted mixed $l_1/l_p$-norms to induce the group-sparsity, in which no prior information of the unknown signal
is assumed other than the fact that it is sufficiently sparse.

In~\cite{IV:yu2013}, considering the
network resources including the fronthaul capacities and the transmit powers at RRHs,
an optimal joint clustering and beamforming design problem is
addressed, in which each user dynamically forms a sparse network-wide
beamforming vector whose non-zero entries correspond to the serving RRHs.
By adopting a compressive sensing approach of using reweighted $l_1$-norm
to approximate the $l_0$-norm, the authors turn the original non-convex
problem into a series of convex weighted power minimization
problem, further solved by a low-complex duality approach.
Simulation results show that the proposed algorithm can achieve a
better tradeoff between the total transmit power and the sum fronthaul
capacity than the existing methods in the high SINR regime.

As a further enhancement of~\cite{IV:yu2013}, which mainly considers the sum power
and sum fronthaul capacity, the authors of~\cite{IV:yu2014}~\cite{IV:yu2015} focus
on network utility maximization under per-RRH transmit
power and the per-RRH fronthaul capacity constraints. Among the family of utility functions,
the weighted sum rate (WSR) has been widely applied to the network control
and optimization problems. Motivated by the idea in
compressive sensing literatures, non-convex $l_0$-norm in the fronthaul constraint
is approximated by the convex reweighted $l_1$-norm. Furthermore, the WSR problem can
be reformulated as an equivalent weighted minimum mean square error (WMMSE)
minimization problem in order to reach a local optimum solution.
To reduce the computational complexity of the proposed algorithm, it is proposed
to iteratively remove RRHs
with negligible transmit power out of the candidate cluster
for each UE, and remove those UEs with negligible rates
out of the scheduling pool.

On the basis of~\cite{IV:yu2014},~\cite{IV:yu2015}
considered the static clustering which fixes the RRH cluster for each UE and
jointly optimized the scheduling and beamforming vectors. Two
static clustering formation algorithms are proposed, namely
the maximum loading based static clustering and the biased signal
strength based static clustering, which both effectively
take the traffic load of each RRH and the
channel condition of each UE into account.
A novel point in~\cite{IV:yu2015} is to
incorporate the per-RRH fronthaul constraint into the
optimization framework. With explicit per-RRH fronthaul constraints,
it is shown that the fronthaul resources can be efficiently
utilized and the network utility can be significantly
improved.

A framework is proposed to design a green C-RAN in~\cite{IV:pre10}, which is formulated
as a joint RRH selection and power minimization beamforming
problem. Considering that the
power consumption of fronthaul is determined by the set of active RRHs,
a group sparse beamforming method is proposed by inducing the
group-sparsity of beamformers using the weighted $l_1/l_2$-norm.
The proposed problem is a mixed-integer non-linear programming problem, which is NP-hard. To solve this problem, two
group sparse beamforming (GSBF) formulation algorithms with different complexities are
proposed, namely, bi-section GSBF and iterative GSBF. It is demonstrated
that the GSBF framework is effective to provide near-optimal solutions to this problem.
The proposed bi-section GSBF algorithm is proved to be a better option for a large-scale C-RAN due to
its low complexity, while the iterative GSBF algorithm can be applied to provide better
performances in a medium-size C-RAN.

\subsection{Lessons Learned}

One of the key advantages in C-RANs is to provide a large-scale centralized signal processing across multiple RRHs and UEs in both uplink and downlink. The feasibility and performance gains of LSCP depends strictly on CSIs. If the capacity of fronthaul is unlimited, the well-known linear beamforming and successive interference cancellation can be directly applied in uplink, and the minimum mean-square error or zero-forcing beamforming strategies can be used to achieve a good performance gain. On the other hand, a broadcast channel is formed and the linear precoding technique can be directly used in downlink C-RANs when the capacity of fornthaul is unlimited.

Since the fronthaul is often capacity constrained, LSCP and compression process should be jointly designed, and the core challenge for LSCP in C-RANs is to design the efficient algorithms when CSIs are non-ideal and the fronthaul is constrained. As an initial LSCP strategy in uplink C-RANs, fixed-rate scaler uniform quantization is preferred when the practical LSCP is designed. In downnlink, the zero-forcing beamforming strategy with scalar fixed quantization can be first used. Furthermroe, substantial overhead to obtain full CSIs for precoding design can be reduced by compressive CSIs acquisition and
stochastic beamforming \cite{IV:SCB1}. With the guidance of the sparsity pattern of the precoding matrix, the complexity of RRH selection can be significantly decreased \cite{IV:MCP}. Finally, per RRH fronthaul constraints with $\mathop l\nolimits_0 $-norm can be handled by iteratively approximating the per RRH fronthaul constraints through a reweighted $\mathop l\nolimits_1 $-norm technique \cite{IV:yu2014}.

\section{Channel Estimation and Training Design}

In C-RANs, the lack of full CSIs presents the main hurdle to the
feasible implementation of C-RANs since an accurate acquisition of instantaneous
CSIs is essential for most of advanced technologies, e.g., interference
coordination and coherent reception. The assumption that the full CSIs are
known at the BBU pool is also not practical due to the time-varying nature of the
radio channel. Consequently, developing accurate and efficient
channel estimation techniques is vital to achieve the performance
improvement of C-RANs \cite{CE0}. In this section, an overview of channel estimation
techniques proposed for C-RANs is provided to give useful guidelines
on the new requirements when employing the conventional channel
estimation approaches in C-RANs. Moreover,
some challenging issues about channel estimation in the fronthaul constrained C-RANs are presented as
the meaningful research topics in the future.

Channel estimation as a fundamental problem in wireless communication systems
has been extensively investigated in the past decades \cite{CE}. In general, channel
estimation can be basically categorized into two kinds of strategies, i.e.,
training-based and non-training-based channel estimation. The training-based channel
estimation is the most common method in the CSIs acquisition of wireless communication
systems due to its flexibility and low complexity. By periodically inserting the training
sequences into data frames in time or frequency domain, the
receiver can easily estimate the CSIs of the radio channels by a prior knowledge of the
training sequences \cite{CE:1}. However, the use of the training sequences degrades
the system performance in terms of SE and EE as additional spectrum resources are demanded
for the training transmission. Therefore, the non-training-based channel estimation
is introduced as an alternative way to obtain the CSIs by adequately using the intrinsic
characteristics of data signal itself instead of pilot signal \cite{CE:2}. Despite the high
SE, there still exits some restrictions of non-training-based channel
estimation, e.g., slow convergence speed, high computation complexity, and sensitivity to
noise, which limit the implementation in practice. In the next, as shown in Table \ref{es}, the
training-based channel estimation in C-RANs is mainly discussed as well as the applications
of non-training-based channel estimation in C-RANs.

\begin{table*}[!htp]
\caption{SUMMARY OF CHANNEL ESTIMATION TECHNIQUES IN C-RANS}\label{es}
\scriptsize 
\centering
\begin{tabular}{|p{0.7 in}|p{1.5 in}|p{1.8 in}|p{1.8 in}|}
 \hline
 \textbf{Literature}& \textbf{Techniques} &\textbf{Approaches}& \textbf{Feature} \\ \hline
 \cite{CE:1} & Training-based channel estimation &Training sequences embedded into data frames and CSIs acquisition based
 on a prior knowledge of training & Easy deployment, high accuracy, technical maturity and low
 spectral efficiency \\ \hline
 \cite{CE:4}, \cite{CE:5}, \cite{CE:6} & Superimposed training based channel estimation & Training sequences superimposed on the data symbols & High resource utilization, poor estimation performance and low received signal-to-noise ratio \\ \hline
 \cite{b1}, \cite{b3} & Segment training based channel estimation & Two consecutive segments to obtain the individual
 CSIs & High estimation accuracy, low spectral efficiency and data transmission rate guarantee \\ \hline
 \cite{CE:2} & Blind/semi-blind channel estimation & Channel estimated without training by using the statistical
 and other properties of the noise and data & High spectral efficiency, long observation time and
 high computational complexity \\ \hline
\end{tabular}
\end{table*}

\subsection{Superimposed Training Based Channel Estimation in C-RANs}

The data transmission in C-RANs is performed by means of two communication links, the
first part by the radio access link (ACL) between the UE and the RRH, and the remaining
part by the fronthaul link connecting the RRH and the BBU pool. The fronthaul links are
wire/wireless, and the wireless fronthaul links (WFLs) are mainly discussed herein due to its
low expenditures and flexible deployment for RRHs. To achieve the optimal
system design in C-RANs, the individual CSIs of these two links should be obtained at the BBU pool
as required by certain technologies, e.g., beamforming design \cite{CE:3}. Under the circumstances,
the traditional training design is inefficient for fulfilling the requirements due to the
high training overhead and the significant performance degradation resulting from the
CSIs feedback during the individual CSIs acquisition. To tackle this problem, the superimposed
training scheme in \cite{CE:4} is a promising solution to the channel estimation
in C-RANs, in which the RRH superimposes its own training sequence on the received one
such that the individual CSIs can be estimated at the BBU pool. In this way, a better tradeoff
between the channel estimation accuracy and the training overhead can be achieved in the superimposed
training scheme with no occupation of additional training resources compared with the
traditional training scheme.

\begin{figure}[t]
\center
\includegraphics[scale=0.35]{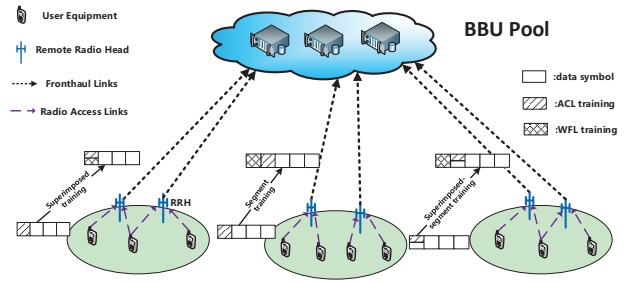}
\caption{Training scheme in C-RANs.} \label{supsegmodel}
\end{figure}

The discussion on the superimposed training based channel estimation can first start
with channel estimation in multiple access relay networks (MARNs) \cite{CE:5}, which
can be extended to C-RANs. As shown in Fig. \ref{supsegmodel}, the superimposed training scheme in \cite{CE:5} considers two phase.
In the superimposed training scheme,
the sources transmit the training matrix $\mathbf{T}$ in the first phase and
each RRH superimposes a training sequence $\mathbf{t}_{r}$ over the received signal in
the second phase. By using these two training signals, the destination node can
separately obtain the CSIs of the composite source-RRH-destination and the individual
RRH-destination links. Based on the superimposed training scheme, the channel estimation
in uplink network coded MARNs is investigated by employing the maximum likelihood (ML) method.
Furthermore, it is pointed out that the optimal training structures of $\mathbf{t}_{r}$ and
$\mathbf{T}$ are orthogonal. The optimal power allocation strategy can be derived by judging the
discriminant of the polynomial. To explore the auto-correlation nature of time-selective fading channel,
a correlation-based iterative channel estimation algorithm in \cite{CE:5} can further improve the
estimation performance in MARNs by observing the instantaneous training signals. It is proved that
the iterative channel estimation algorithm is convergent and can provide better accuracy than the ML method.

\subsection{Segment-Training Channel Estimation}

In the superimposed training scheme, the data transmission rate is still not high and
the accuracy of channel estimation is not efficient due to the fact that training sequences
and data symbols have to share the transmitted power. On the other hand, the performance of
channel estimation is greatly influenced by the unknown data symbols when extracting the
training sequences from the compounded signals, and it is difficult to apply in practice
due to the sensitivity to distortion. The segment training scheme in \cite{b1}
can be applied to acquire the individual CSIs of the ACLs and the WFLs in C-RANs, which is an effective way to
improve the estimation accuracy. In the segment training scheme, two consecutive
segments as illustrated in Fig. \ref{supsegmodel} are implemented for the CSIs acquisition of ACLs and WFLs,
respectively. At the beginning, the CSIs of the WFLs can be obtained at the BBU pool by using the first segment
training. Then, based on the estimated CSIs of the WFLs, the CSIs of the ACLs is readily acquired by employing
the training-based channel estimation method. Generally, the segment training scheme can achieve higher estimation
accuracy as the power originally shared by the training and data is allocated to the training alone. The segment
training based channel estimation can be recognized as an alternative for the acquisition of CSIs in C-RANs when
a high estimation accuracy is required.

\subsubsection{Balance between Performance and Overhead}

It should be noted that straightforward implementation of the segment training scheme in C-RANs
would degrade SE since the number of symbols assigned for the training is
sufficiently large. A superimposed-segment (SS) training scheme in \cite{CE:6} can be
regarded as a good way to achieve the balance between the data rate and training overhead in
C-RANs. In the SS training scheme, the superimposed training $\tilde{\mathbf{t}}_{r}$ is
implemented for the ACLs to reduce the training overhead, while another individual segment is
assigned for the training $\mathbf{t}_{s}$ of the WFLs to avoid the data rate reduction, which is
shown in Fig. \ref{supsegmodel}. Due to the time-varying characteristics of the ACLs, the complex-exponential
basis-expansion-model is adopted to model the ACLs. To obtain the individual CSIs of ACLs and WFLs,
the maximum a posteriori probability channel estimation algorithm can be applied. By iteratively
calculating the estimates of ACLs and WFLs, the channel estimation accuracy can be
further improved. Moreover, the iterative channel estimation algorithm is convergent, and the
channel estimation accuracy is higher than the ML method.

\subsubsection{Estimation in Large-scale C-RANs}

In the large-scale C-RANs, the channel estimation is still challenging
since there are lots of channel parameters to be estimated simultaneously.
As a result, the computational complexity is another important issue for the channel estimation in
C-RANs. The conventional linear minimum-mean-square-error (LMMSE) estimator has a cubic
complexity in the dimension of the covariance matrices due to the matrix inversion, and it becomes
awfully large to be unaffordable in the large-scale C-RANs. For the complexity of MMSE estimator,
\cite{b3} has successfully reduce it to a square complexity by applying the matrix polynomial expansion (PE).
PE is a well-known technique to reduce the complexity of large-dimensional matrix inversions
by approximating a matrix function by an $L$-degree matrix polynomial. Inspired by the low computational
complexity estimator, the same method can be applied to the channel estimation in C-RANs with segment
training scheme. First, the LMMSE estimator is employed to obtain the CSIs of the ACLs and WFLs. When
implementing the LMMSE estimator, the matrix inversion of the covariance matrix of the received training
can be replaced by the PE. In this way, the computational complexity of the segment training based
channel estimation is significantly reduced.

\subsection{Non-Training-Based Channel Estimation in C-RANs}

The training-based channel estimation has demonstrated its effectiveness to
acquire a reliable estimate of the CSIs in C-RANs. However, the training overhead of
training-based channel estimation is relatively significant, and the transmission bandwidth
efficiency is decreased in turn. The training is undesirable for
some special scenarios, and the training-based channel estimation has not
exploited the intrinsic properties of data symbols. Thus, the non-training-based
channel estimation can be employed to resolve the problem of spectrum resource
scarcity \cite{CE:7}. However, a large number of data symbols is
required to improve the channel estimation accuracy blindly. The inherent ambiguity, e.g., phase
ambiguity, also blocks the implementation of the blind channel estimation in C-RANs if merely
relying on the data samples to estimate the CSIs. Still, a small number of training is needed
to obtain \textcolor{red}{an} initial estimate.
Semi-blind channel estimation as a hybrid of blind and training based
approaches is considered as a pragmatic compromise to take their strong points to offset
the weakness. By incorporating both the training signal and the data samples, the semi-blind
channel estimation can achieve higher accuracy than purely using the training-based or blind
estimation. Moreover, the study of the semi-blind channel estimation in C-RANs is in its initial
phase, and there are only tentative attempts at the related research area.

Inspired by the work in \cite{CE:9}, the semi-blind channel estimation is employed in C-RANs to
achieve a good accurate estimation without sacrificing the bandwidth efficiency. The least-squared
estimation is used to obtain an initial estimate, and the quasi-Newton method is applied
to further improve the estimation accuracy based on the maximum likelihood principle. By
treating data symbols as Gaussian-distributed nuisance parameters, the joint likelihood function
of both the training and data can be derived. The maximization problem of the joint likelihood
function is actually a nonlinear problem, and an iterative quasi-Newton method can be applied to
obtain the CSIs in C-RANs with a limited number of data symbols. In this way,
the semi-blind channel estimation can achieve substantial improvements over the conventional training-based method in
terms of estimation accuracy.

\subsection{Lessons Learned}

Compared with the traditional training design scheme, a good tradeoff between the channel estimation accuracy and the training overhead
can be achieved in the superimposed training scheme for C-RANs. Meanwhile, the non-training-based channel estimation can achieve high SE but meanwhile with
high computing complexity. Accurately, by incorporating both the training signal and the data
samples, the semi-blind channel estimation can achieve higher accuracy than the pure training-based or blind estimation schemes.

Furthermore, since the constrained fronthaul is one of the major practical hurdles for C-RANs,
the channel estimation by taking fronthaul loading into consideration is an
important and challenging issue, which needs to be urgently tackled in the future. The channel estimation errors should be taken into consideration and the
corresponding influence is urgent to be exploited.

\section{Cooperative Radio Resource Allocation}

To maximize SE and EE in C-RANs,
the multi-dimensional radio resources should be optimized,
which often leads to non-convex problems.
When the fast adaptation to users' traffic is required,
cross-layer optimization needs to be tackled. In this section, selected published works
, related to the static CRRA without QSI and the dynamic CRRA with queue-awareness,
are discussed and summarized.

\subsection{Static CRRA without QSI}

For the static CRRA without QSI, the classic methods like Lagrange dual method
and WMMSE method can be directly adopted. In addition, to alleviate the transmission burden on the fronthaul,
the game model based optimization approaches can be used as well.

\subsubsection{Classic Non-convex Optimization}

\begin{figure}[!htp]
\center
\includegraphics[scale=0.55]{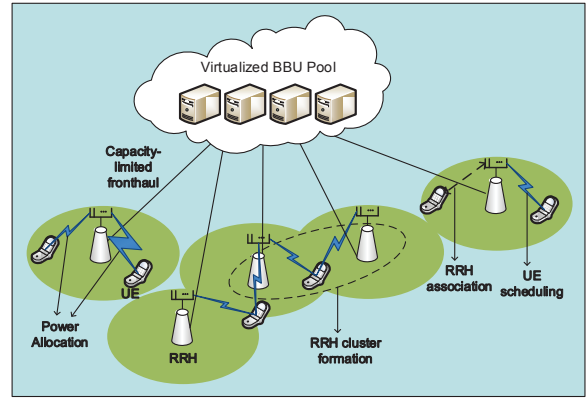}
\caption{Multi-dimensional radio resource allocation in C-RANs.} \label{resource}
\end{figure}

The performances of C-RANs, such as SE and EE, are mainly determined by the multi-dimensional CRRA, including the power allocation or precoding design, user scheduling, RRH cluster formation, and so on, as shown in Fig. \ref{resource}. To achieve the global optimal solutions to the static CRRA problems without QSI, centralized methods are effective because the multi-dimensional CRRA which influences the system performance can be monitored by the BBU and optimized jointly. These centralized method have attracted much attention recently \cite{r8,n2,n3,n4,n5}. However, because of the involvement of integer variables like RRH selection indicator and the existence of inter-tier and intra-tier interference, the joint optimization of multi-dimensional CRRA is always non-convex, making the optimization problem hard to solve. To handle the non-convexity, in general, there are two kinds of centralized methods. One is using heuristic algorithms which are suitable for finding the optimal solutions to NP-hard problems. The other one is to transform these non-convex problems into convex problems by different methods, and then the primal problems can be solved by settling the corresponding convex problems. For the application of heuristic algorithms, in \cite{1}, the author proposed a branch and bound based algorithm to optimize the system power consumption with the constraints of transmit delay, transmit power and fronthaul capacity.
Although a global optimal solution to the problem can be achieved, the complexity of the algorithm is sufficiently high. Hence, in the following, we mainly focus on the centralized methods that can transform non-convex problems into convex problems with a low complexity.

Firstly, Lagrange dual method is effective to design centralized algorithms that can optimize resource allocation in C-RANs when the primal problems satisfy time-sharing condition. In \cite{2}, the author investigated the EE maximization for the heterogeneous cloud radio access network. Due to the non-convexity of the primal problem, the author tackled the primal problem by solving its dual problem aiming to minimize the Lagrange function of the primal problem, and the dual problem is always convex by definition. Moreover, since the primal problem satisfies the time-sharing condition, zero duality gap holds. Thus, the non-convex primal problem can be equivalently transformed into a convex optimization problem, i.e., its dual problem.

Secondly, WMMSE has been widely applied to design the centralized power allocation in C-RANs \cite{1,3,IV:yu2015,5}. Due to the potential intra-tier interference and limited cluster size, the expression of SINRs is no longer convex with regard to the entries of the precoding matrix, causing the non-convexity of the whole problem. Fortunately, WMMSE has been proved to be effective in transforming this kind of non-convex optimization problems into convex optimization problems, which can achieve near optimal solutions.

Specifically, in \cite{1}, besides the Branch and Bound method, the author adopted WMMSE to transform the non-convex power consumption minimization problem into
a convex optimization problem with respect to the entries of the precoding matrix. A WMMSE based precoding scheme to maximize the downlink system throughput is proposed in \cite{3,IV:yu2015}. In \cite{3}, the joint power and antenna selection optimization in a C-RAN is considered. The throughput optimization problem is decomposed into two subproblems. One of the subproblems, which is non-convex, can be transformed into a convex optimization problem with respect to the entries of the precoding matrix by WMMSE. In \cite{IV:yu2015}, the authors optimize the weighted throughput by the optimal user scheduling, RRH clustering, and precoding design scheme. Due to the intra-tier interference, the objective function is non-convex, and it is transformed into a convex form by WMMSE. While in \cite{5}, the uplink transmission is taken into consideration and a joint downlink and uplink user-RRH association and precoding scheme is proposed to minimize the power consumption, in which the joint downlink and uplink optimization problem is transformed into an equivalent downlink problem. As a result, WMMSE can be used to transform the non-convex downlink problem into a convex problem with respect to the entries of the precoding matrix.

Thirdly, for some precoding design problems in C-RANs, $l_0$-norm is incorporated to express RRH selection, which leads corresponding problems to integer programming problems. Moreover, it should be noted that precoding matrices in C-RANs are sparse because the number of RRHs is much more than the number of users, which indicates that the percentage of the non-zero elements in precoding matrice is low. To transform such non-convex precoding design problems into convex problems, $l_1$-norm approximation method can be used due to the excellent convex approximation to $l_0$-norm and the convenience to induce the sparsity of precoding matrices. In \cite{1}, the $l_1$-norm approximation is utilized to transform the non-convex $l_0$-norm objective function into a convex form. In \cite{IV:yu2015}, the author investigated the reweighted $l_1$-norm approximation to deal with the $l_0$-norm in the constraint and proposed an iterative weight calculation scheme.

In the $l_1$-norm approximation method, each coefficient in the precoding matrix is independent, however, in C-RANs, such independence does not always hold. For example, in C-RANs, one user is always served by a selected cluster of RRHs, which illustrates that the elements not belong to these RRHs in precoding matrix are set to zero \cite{5}. Besides, one RRH can be switched off when all of its coefficients in the precoding matrix is set to zero \cite{IV:pre10}. In these cases, the coefficients of precoding matrices should be optimized jointly rather than individually, and thus the $l_1$-norm approximation can't be used directly because the zero entries of the precoding matrices may not align in the same RRH.
To overcome this challenge, mixed $l_1/l_p$-norm approximation method can be adopted to induce group sparsity. In \cite{IV:pre10,5}, mixed $l_1/l_p$-norm approximation method is proposed to handle a group sparse based RRH selection problem. The $l_1/l_p$-norm means the $l_1$-norm of an $l_p$-norm. When the mixed $l_1/l_p$-norm approximation method is applied in C-RANs, the coefficients of the precoding matrix will be divided into different groups which can be described as the $l_p$-norm. Then execute the traditional $l_1$-norm approximation with respect to these $l_p$-norm groups and the mixed $l_1/l_p$-norm approximation is completed. In \cite{5}, traditional mixed $l_1/l_p$-norm method is used to transform group sparse based $l_0$-norm constraints. In \cite{IV:pre10}, a three stage group sparse precoding design algorithm is proposed. The non-convex $l_0$-norm constraints are transformed into convex forms by weighted mixed $l_1/l_p$-norm method. Further more, the author described the way to design the weights for different coefficient groups to achieve significant performance gain.

In fact, the centralized methods mentioned above are not independent from each other. When solving a non-convex problem, one or more different methods may be utilized. For example, in C-RANs, the precoding matrix design and the RRH selection problem may be optimized at the same time. Thus, the WMMSE method and the norm approximation method can be jointly used to transform the non-convex primal problem into a convex problem. In \cite{IV:yu2015}, the non-convex primal problem is transformed into the well known quadratically constrained quadratic programing problem by the WMMSE method and the norm approximation method. The QCQP problem can then be easily solved by some classic methods such as CVX. In \cite{3,n6}, the author transformed the non-convex primal problem into a convex problem and then used the lagrange dual method to transform the convex problem into a more tractable form.
Centralized methods to optimize resource allocation in C-RANs are summarized in Table \ref{centralized}.

\begin{table*}[!htp]
\caption{SUMMARY OF CENTRALIZED METHODS TO OPTIMIZE RESOURCE ALLOCATION IN C-RANS}\label{centralized}
\scriptsize 
\centering
\begin{tabular}{|p{0.5 in}|p{1.1 in}|p{2.8 in}|p{1.4 in}|}
 \hline
 \textbf{Literature}& \textbf{Method} &\textbf{Suitable Problems}& \textbf{Feature} \\ \hline
 \cite{1} & Branch and Bound method &Suitable for most kinds of resource allocation problems in C-RANs & Optimal solution but with high complexity \\ \hline
 \cite{2} & Lagrange dual method & Suitable for problems with zero duality gap & Optimal solution \\ \hline
 \cite{IV:yu2015,1,3,5}& WMMSE & Suitable for problems whose non-convexity is caused by the inter-tier or intra-tier interference & Sub-optimal solution \\ \hline
 \cite{IV:pre10,IV:yu2015,1,5} & Norm approximation & Suitable for problems involving integer variables and being equivalent to a $l_0$-norm & Sub-optimal solution \\ \hline
\end{tabular}
\end{table*}

\subsubsection{Game Model based Optimization}

The centralized radio resource allocation optimization discussed above requires the availability of full and ideal CSIs at the BBU pool, which imposes a significant burden on the fronthaul, especially for large-scale C-RANs. To alleviate this burden and decrease the computing complexity, one may resort to decentralized solutions whereby RRHs self-organize the radio resource allocation and make decisions in a situation of interest conflict based only on collected local CSIs. To this end, the framework of game models can be adopted to allocate the radio resources in C-RANs. Detailed
investigations of game models in wireless networks have also been surveyed in \cite{gamemodel}, but here we focus more on the opportunities of using the game models for CRRA in C-RANs.
As illustrated in Fig. \ref{differentgame}, two kinds of game models, e.g., coalition and contract, have been applied in C-RANs.

For the coalition game model, it is adopted to partition RRHs of C-RANs like BSs in traditional cellular networks into different clusters called coalitions. In each coalition,
RRHs are cooperative. Specifically, in uplink, authors in \cite{VI-3-1} made multiple single
antenna transmitters self-organize into multiple coalitions, where transmitters in the same coalition form a virtual
multi-antenna array. In downlink, RRHs can distributively form different coalitions, and
interference alignment \cite{VI-3-2} or beamforming \cite{VI-3-3} can be performed in each coalition to
mitigate inter-cell interference.

In C-RANs, RRHs form different clusters to serve UEs, and identifying proper
schemes to form clusters is essential to give a high performance of C-RANs.
Although the BBU pool enables centralized optimization, which brings much performance gain,
a heavy burden
can be imposed on the fronthaul links to obtain global CSIs, especially for large-scale C-RANs.
To reduce transmission burden
on the fronthaul links, it is expected to involve game model based optimization schemes, where RRHs selfishly optimize their own
objectives only based on local information. In \cite{VI-3-4}, stochastic geometry is
combined with coalition formation game. First,
with a fixed intra-cluster cooperation strategy,
authors derived an explicit expression for the successful
access probability. Then, a utility function is defined, which contains the benefit as well as the cost incurred by clustering.
After that, the problem of RRH clustering
is modeled as a coalition formation game, and a merge-split algorithm is
used to cluster RRHs. Cases with and without size limitation of RRH clusters
are both considered. By simulations, it can be seen that
the proposed merge-split based RRH cluster formation algorithm
always outperforms no
clustering strategy since interference is reduced with clustering.
Moreover, the impact of the parameter
reflecting the cluster formation cost is evaluated. It is found that
a lower value of the parameter, which means a lower cost for a cluster with fixed size,
leads to better performance. This is because
lower cluster
formation cost leads to larger cluster size, and thus more interference
can be eliminated.

For the contract game model, it is mainly used to give nodes in wireless networks incentive to reach an agreement
under asymmetric information scenario in H-CRANs. The BBU pool offering the MBS a contract is called a principle, and the MBS
that will potentially accept the contract is called agent. In literatures, contract game model has been applied to
spectrum sharing and relay selection. Specifically, to solve spectrum sharing problem in cognitive
radio networks, a resource-exchange based contract model and a money-exchange based contract model are adopted in
\cite{VI-3-41} and \cite{VI-3-42}, respectively, with the exact private information about secondary UEs unknown to the spectrum owner.
In \cite{VI-3-43}, also under an asymmetric information scenario where the source is not fully informed about the potential relays,
a contract is designed to help the source select relays to optimize its own utility.

When the C-RAN is underlaid with MBSs in H-CRANs, mutual interference between the C-RAN and MBSs
should be coordinated. Considering the C-RAN and MBSs hold an equal position,
it is preferable to implement distributed interference coordination.
To this end, \cite{VI-3-5} proposed a contract based interference coordination approach for downlink
scenario containing a C-RAN and an MBS.
The BBU pool is taken as a principle and the MBS is taken as an agent.
The proposed approach depends on a cooperative strategy, where three phases are involved in
one time transmission
interval with a fixed time length ${T_0}$.
In the first phase with time length ${t_1}$, RRHs serve all RRH UEs (RUEs) and the macro UE (MUE), which is named as \emph{RRH-alone with UEs-all scheme}.
In the second phase with time length ${t_2}$, all RUEs are served by RRHs, and the MUE is not served neither by the MBS nor RRHs.
This scheme is called \emph{RRH-alone with RUEs-only} scheme. In the last phase,
\emph{RRH-MBS with UEs-separated}
scheme is performed, where RRHs and the MBS serve RUEs and the MUE, respectively.
To improve the system performance with the proposed strategy, ${t_1}$ and ${t_2}$ should
be optimized, aiming at maximizing the utility of the BBU pool.
Particularly, when the BBU pool knows the exact CSIs of the MBS-MUE link, individual rational constraints should be met, and
when the BBU pool only knows the statistical information about the MBS-MUE link, both of individual rational constraints and incentive compatible constraints should be met.
Via simulations, the performance gain of the proposed approach compared to FRPC approach and TDIC approach is verified.
For the FRPC, the same resource block is reused by the C-RAN and the MBS, and the transmit powers of RRHs and the MBS
are determined according to the fairness power control. For the TDIC,
the C-RAN and
the MBS transmit separately in different time phases.

Besides the coalition formation game and the contract game model, the matching game model, a Nobel-prize winning framework, is promising
as well, since it provides mathematically tractable solutions
for the combinatorial problem of matching players in two distinct sets, depending on the individual
information and preference of each player \cite{VI-3-6}.
However, the application of matching game model to C-RANs is still unexplored.
One possible application is to use matching game model to solve the RRH association problem.
While many works have applied the matching game model to BS association problems \cite{VI-3-7,VI-3-8,VI-3-9}. Each UE can only be associated with one BS in these works,
which is definitely different from the situation in C-RANs. In C-RANs, each UE can be associated with multiple RRHs simultaneously,
making the RRH association problem hard to be solved. As a result, how to use the matching game model to develop a low-complexity
RRH association algorithm for C-RANs is still not straightforward and should be researched in the future. A summary of literatures applying game models to C-RANs is shown in Table \ref{game model}, and
a summary of advanced game models is shown in Table \ref{game model1}.

\begin{table*}[!htp]
\caption{SUMMARY OF LITERATURES APPLYING GAME MODELS TO CLOUD-RANS}\label{game model}
\scriptsize 
\centering
\begin{tabular}{|p{0.5 in}|p{0.8 in}|p{2.2 in}|p{2.3 in}|}
 \hline
 \textbf{Literature} & \textbf{Game model}& \textbf{Contribution} &\textbf{Simulation result}\\ \hline
 \cite{VI-3-4} & Coalition formation game& Coalition formation game is applied to RRH cluster formation, and merge-split
 algorithm is used to give a solution &The performance of clustering strategy is always better than no
clustering strategy, and lower cluster
formation cost leads to better system performance \\ \hline
 \cite{VI-3-5} & Contract theory & Contract theory is applied to interference coordination between the C-RAN and the MBS, and optimal contract
 design is handled under perfect CSIs as well as imperfect CSIs & Significant performance gain is achieved by the proposed approach compared to
 FRPC and TDIC schemes\\ \hline
\end{tabular}
\end{table*}

\begin{table*}[!htp]
\caption{SUMMARY OF ADVANCED GAME MODELS}\label{game model1}
\scriptsize 
\centering
\begin{tabular}{|p{0.8 in}|p{2 in}|p{1.5 in}|p{1.5 in}|}
 \hline
 \textbf{Game model} & \textbf{Feature}& \textbf{Advantage} &\textbf{Application}\\ \hline
 Coalition formation game & A set of players form cooperative groups, i.e., coalitions, in order to strengthen their positions in a given situation& Suitable for grouping players of the same category with identical utility functions & Inter-cell interference mitigation \cite{VI-3-2}\cite{VI-3-3}, cluster formation \cite{VI-3-4} \\ \hline
 Contract theory & The principle offers agents a contract without knowing complete agents' private information, aiming at maximizing its own utility & Suitable for handling cooperation under asymmetric information scenario & Spectrum sharing \cite{VI-3-41}\cite{VI-3-42}, relay selection \cite{VI-3-43}, interference coordination \cite{VI-3-5}\\ \hline
 Matching theory & The players are divided into two sets, and the main goal of matching is to optimally match players in one set with players in the other set, given all players' individual objectives and learned information & Suitable for solving the combinatorial problem of matching players in two distinct sets & BS association \cite{VI-3-7}\cite{VI-3-8}\cite{VI-3-9}\\ \hline
\end{tabular}
\end{table*}

\subsection{Dynamic CRRA with Queue-Awareness}

In C-RANs, all RRHs are connected to a BBU pool via fronthaul
links and centrally managed by the BBU pool. Due to an increased
level of frequency sharing and RRH density, the inter-RRH
interferences should be elaborately controlled through advanced
cooperative signal processing techniques in physical (PHY) layer. In
addition, the centralized C-RANs should be fast adaptive to the
changes of traffic demands caused by users' joining, leaving, or
bursty traffic requests at any time. Therefore, to maximize the
performance from a global perspective, the cross-layer optimization
framework for cooperative radio resource management (RRM) is
essential for future C-RANs. The concept of cross-layer
optimization is not new \cite{crosslayer1}, and solving
this kind of optimization problems is known to be computationally
complex. However, with the cloud-computing enabled C-RANs, the
time has come to consider the use of advanced cooperative signal
processing techniques and efficient cooperative RRM algorithms for
network provision. The cross-layer resource optimization in
C-RANs integrates several important cross-layer techniques: 1)
advanced cooperative signal processing techniques for PHY layer
interference management; 2) medium access control (MAC) layer
algorithms to handle user scheduling, time-frequency resource
allocation, RRH clustering, e.t.c; 3) network layer solutions to
control individual traffic flows and to manage the network services.
Such cross-layer optimization is more challenging than the
traditional cellular network optimization as it involves complicated
interactions across different layers.

Lots of articles on cross-layer resource optimization in wireless
systems have been published recently, most of which have illustrated
that significant SE or energy efficiency EE
gain can be obtained by the cross-layer optimization of radio
resources. The centralized architecture of C-RANs facilitates
the implementation of CoMP transmission
and reception to improve the overall SE. In this context, the
cross-layer design issue of RRH clustering and beamforming to
strike the best tradeoff between system throughput and signaling
overhead has been systematically studied in \cite{IV:yu2015}.
The cross-layer optimization of joint user scheduling and
beamforming has been investigated in \cite{ZQLuo}
and \cite{WYu}, where the cross-layer resource
optimization is conducted in a way that maximizes a system utility
including throughput gain and proportional fairness. The
densification of RRHs in C-RANs gives users the flexibility of
choosing their serving RRHs from a large number of nearby RRHs. This
freedom of RRH selection is properly exploited
in \cite{3}, which achieves substantial gains in
terms of both network throughput and user fairness. A unified way
called WMMSE algorithmic
framework to design various cross-layer resource optimization
algorithms is proposed in \cite{HBaligh}, which can be
implemented in a parallel manner.

\begin{figure}
\centering
\includegraphics[scale=0.45]{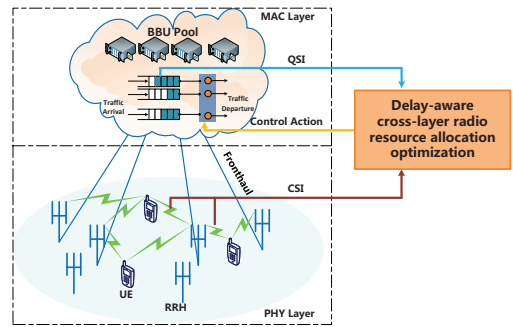}
\caption{Delay-aware cross-layer radio resource allocation
optimization in C-RANs} \label{fig_delay_aware}
\end{figure}

However, existing literatures \cite{IV:yu2015,3,ZQLuo,WYu,HBaligh}
are typically based on the assumption of infinite queue backlogs and
stationary channel conditions. Therefore, only some performance metrics including the SE, EE, proportional fairness are optimized, and the resulting control policy
is only adaptive to CSIs. Nevertheless,
since the practical C-RANs often operate in the presence of
time-varying channel conditions and random traffic arrivals, and the
CSIs-only cross-layer designs cannot guarantee good delay
performance. As shown in Fig. \ref{fig_delay_aware}, the dynamic
cross-layer optimization algorithms should be adaptive to both CSIs
and QSI. A delay-aware cross-layer
optimization framework, taking both the traffic queue delay and PHY
layer performance into account, is not trivial, as it involves both
queuing theory (to model the traffic queue dynamics) and information
theory (to model dynamics in PHY). It is made more
complicated by practical issues in C-RANs including fronthaul
capacity limitations, system states (including QSI and CSIs)
overhead, and parallel implementation of the algorithms.

Some developments in this area have been made, and the potential
approaches with their advantages and disadvantages have been
identified. The design approaches towards delay-aware cross-layer
optimization basically come down to three types: the equivalent rate
approach, the Lyapunov optimization approach and the Markov decision
process (MDP) approach. The following surveyed references are by no
means exhaustive. Nonetheless, they still help to present the main
roadmap of ongoing attempts to design delay-aware cross-layer
optimization algorithms for future C-RANs.

\subsubsection{The Equivalent Rate Approach} The
principle idea is converting the average delay constraints (or
objectives) into equivalent average rate constraints (or
objectives). The equivalence is usually established based on queuing
theory or large deviation theory. In \cite{Zarakovitis},
the relationship among the average delay requirement, the average
arrival rate and the average service rate is established using the
packet flow model of queuing theory, and the cross-layer
optimization problem is solved with the equivalent service rate
constraints in PHY. In \cite{CLin}, the
statistical delay QoS requirement is characterized by the QoS
exponent according to the concept of effective capacity from large
deviation theory, and the distributed cross-layer resource
allocation scheme is proposed to deal with the co-channel
interference and guarantee the individual statistical delay QoS
requirement.

Intuitively, the equivalent rate constraint approach provides
potentially simple solutions for delay-aware cross-layer resource
optimization in C-RANs in the sense that the problem can be
transformed into a pure PHY layer optimization problem, and the
traditional PHY layer optimization approach
in \cite{ZQLuo}-\cite{HBaligh} can be then readily
applied to solve the transformed problem. The optimal control policy
is a function of the CSIs with some weighting shifts by the delay
requirements, and hence it is simple to implement in practical
communication systems. However, due to the complex coupling among
the queues induced by multi-user interference, it is difficult to
express delay constraints in terms of all the control actions.
Therefore, this approach cannot be easily extended to C-RANs
with different levels of interference and fronthaul capacity
constraints.

\subsubsection{The Lyapunov Optimization Approach}

The Lyapunov optimization approach is an extension of Lyapunov drift
theory, which is originally exploited to analyze the characteristics
of the control policies in the stochastic stability
sense \cite{MNeelyBook}. With Lyapunov optimization, we
can stabilize the queues of networks while additionally optimizing
performance metrics or satisfying additional constraints across the
layers. The delay-guaranteed feature of Lyapunov optimization comes
from the fact that the traffic queues are strongly stable if they
have bounded time average backlogs. Using the Lyapunov optimization
theory, the procedures of deriving delay-aware cross-layer
optimization algorithms can be summarized as three main
steps \cite{MNeelyInfocom}: 1) if needed, transform
other average performance constraints into queue stability problems
using the technique of virtual cost queues; 2) choose a Lyapunov
function (usually in quadratic form) and calculate the Lyapunov
drift-plus-penalty function; 3) based on the system state
observations, minimize the upper bound of Lyapunov
drift-plus-penalty function over all polices at each time slot.

The Lyapunov optimization has been successfully used to derive
delay-aware cross-layer optimization algorithms, which provides a
quantitative method to obtain the desired performance tradeoff.
In \cite{JTChen}, a two-stage delay-aware cross-layer
RRM algorithm using the Lyapunov optimization is proposed in a
multi-user multi-input-multi-output (MU-MIMO) systems, which can be
looked as a general model of C-RANs. The proposed cross-layer
RRM algorithm consists of a delay-aware limited CSIs feedback
optimization stage over a long timescale and a user scheduling stage
over a short timescale. The numerical results
in \cite{JTChen} demonstrate a controllable tradeoff
between average amount of CSIs feedback and average queue delay. The
Lyapunov optimization theory also has applications in the network
layer traffic flow control when the exogenous arrival rates are
outside of the network stability region. By incorporating the
congestion control into the radio resource optimization using the
lyapunov optimization technique, the delay-aware cross-layer
algorithm is proposed in \cite{HJuBLiang} to maintain
the delay performance and maximize the throughput utility for LTE-A
networks. The problem of EE-guaranteed throughput-delay tradeoff in
interference-free C-RANs is investigated
in \cite{YZLiT} under a unified Lyapunov optimization
framework, where simulation results demonstrate that the traffic
flow control has the advantages of stabilizing traffic queues and
reducing power consumption.

The Lyapunov optimization approach provides a simple alternative to
deal with delay-aware cross-layer optimization problems. The derived
cross-layer control policies are adaptive to both the CSIs and the
QSI. However, for the Lyapunov optimization theory, the queue
stability is a weak form of delay performance, and the average delay
bounds are tight when the traffic load is sufficiently high.

\subsubsection{The MDP Approach}

In C-RANs, the system state is characterized by the aggregation
of CSIs and QSI, and it evolves as a controlled Markov chain. Thus
the delay-aware cross-layer optimization problem can be modeled as
an infinite horizon average cost MDP, the solution of which can be
derived by solving a Bellman equation. However, the cardinality of
the system state space is exponential w.r.t. the number of queues in
C-RANs and hence solving the MDP is quite complicated in
general. Therefore, the curse of dimensionality, and the
requirements of low complexity and parallel solution are the two
main issues associated with the MDP approach. Furthermore, the
existence of multi-user interference makes the queues of different
UEs couple with each other, making the cross-layer optimization more
complex. Fortunately, the approximate MDP and stochastic learning
methods can help to ensure the low-complexity and parallel
implementation of the delay-aware cross-layer resource
optimization \cite{Borkar}.

To achieve the flexible tradeoff between cooperation gain and
fronthaul consumption, a hybrid CoMP scheme is designed
in \cite{JLiISJ} for the downlink transmission in
C-RANs with capacity-limited fronthaul links and delay-sensitive
traffic. To maintain the low traffic delay, the delay-aware power
and rate allocation with fronthaul capability and RRH power
constraints are then formulated as an infinite horizon constrained
MDP in \cite{JLiISJ}, where the linear approximation,
online learning of value functions and stochastic gradient algorithm
are presented to obtain a low-complexity and parallel solution.
However, the online learning of value functions may suffer from slow
convergence and lack of insight. In \cite{FZhangLCDC},
the delay-aware cooperative beamforming control for delay-sensitive
traffic with limited fronthaul data transfer in C-RANs is
considered, which is also formulated as an infinite horizon
constrained MDP. To tackle the emerging slow convergence of value
functions, the continuous time Bellman equation is derived and the
value function is calculated using the well-established calculus
theory. However, the unavoidable interference brought by imperfect
CSIT results in some loss of strict optimality of value
function \cite{FZhangLCDC}.

The performance degeneration of general MDP-based delay-aware
resource optimization algorithms caused by the suboptimal value
function is still hard to capture and the tradeoff between
optimality and complexity still remains open. It should be mentioned
that one potential method to calculate the optimal value function is
based on the perturbation analysis of value function, as
in \cite{FZhang}. However, the calculation is extremely
complicated and it can only apply to the simple scenario with simple
control actions and lightly coupled traffic queues.

A brief comparison of the three approaches is made
in \cite{YCui}, where three different subcarrier and
power allocation algorithms in an OFDMA system are proposed. The
simulation results demonstrate that the delay performance of the MDP
approach is better than those of the equivalent rate constraint
approach and the Lyapunov optimization approach over a wide
operating regime, while the equivalent rate constraint approach and
the Lyapunov optimization approach have relatively simplified
low-complexity solutions. A summary of the three approaches is
listed in Table \ref{cross}.

\begin{table*}[!htp]
\caption{SUMMARY OF APPROACHES TO DYNAMIC CRRA WITH QUEUE AWARENESS}\label{cross}
\scriptsize 
\centering
\begin{tabular}{|p{0.6 in}|p{0.8 in}|p{2.6 in}|p{1.8 in}|}
 \hline
 \textbf{Literature}& \textbf{Approaches} &\textbf{Solutions}& \textbf{Challenges} \\ \hline
 \cite{Zarakovitis}, \cite{CLin} & Equivalent rate constraint approach &Consider an equivalent problem in PHY by converting
the average delay constraints into average rate constraints & The
solutions are only the functions of CSIs and cannot be easily extend
to C-RANs \\ \hline
 \cite{MNeelyInfocom,MNeelyBook,JTChen,HJuBLiang,YZLiT} & Lyapunov optimization approach & Consider the resource optimization
problem in a stochastic stability sense and obtain an online
solution by minimizing the upper bound of Lyapunov
drift-plus-penalty function & The stability is a weak form of
traffic delay and the derived solutions may not have good
performance in the small delay regime \\ \hline
 \cite{Borkar,JLiISJ,FZhangLCDC,FZhang} & Markov decision process approach & Consider the dynamics of the
system as a controlled Markov chain, and obtain the solution by
solving a Bellman equation together with stochastic leaning or
differential equation & The huge dimensionality of state space and
the requirement of low-complexity solution are the two challenging
issues \\ \hline
\end{tabular}
\end{table*}

\section{Challenges and Open Issues}

In this section, the challenges and benefits brought by the applications of several attractive techniques in C-RANs are discussed,
including edge cache, big data mining, social-aware D2D, CR, SDN, and physical layer security. In addition, the progress of testbed development and trial tests is introduced as well.

\subsection{Edge Cache in C-RANs}

Given the exponential increment of the capacity of modern
storage units and the consistent decline of the cost per stored bit \cite{t-2}, it is possible to endow RRHs with
large storage capabilities at an acceptable cost. Hence, the burden on the fronthaul links can be alleviated
by proactively downloading the data that will be requested with a high
probability and then caching it. Moreover, the storage capabilities of UEs can be utilized to serve other surrounding UEs as well.
In this way, when the requests for the cached data arrive,
they will be served locally, which means the requested data need not be fetched via fronthaul links.
In addition, since the data is more closer to UEs, quality of experience (QoE) can be significantly improved because of
a low transmit delay.

Although the edge cache at edge devices, i.e., RRHs and UEs, brings benefits as mentioned above,
there are several technical challenges to be handled when edge cache is applied to C-RANs.
The first one is the edge cache strategy. Specifically, the frequency of data update should be
judiciously chosen. High frequency update gives UEs better QoE but cost more fronthaul resources, and
low frequency update is the opposite. Meanwhile, since the popularity of
data is the main factor that should be taken into account for the edge cache \cite{t-3},
the edge cache strategy under the incomplete information on the popularity is a challenge.
The second issue is data fetching strategy.
With edge cache involved in C-RANs, the data requested by UEs can now
be found in neighbor edge devices. Then, if the data requested by a UE is not cached in these
devices, an efficient data fetching strategy should be developed to help to decide where to fetch the data and the corresponding route.
The last challenge is the RRH association strategy.
In \cite{t-4,t-5}, some base station association
strategies that consider the impact of edge cache have been developed for small cell networks. Nevertheless, these strategies are designed for the case where
one UE can only be associated with one base station. While, in C-RANs, one UE is often served by multiple RRHs simultaneously.
Consequently, an advanced RRH association strategy taking edge cache into account should be exploited.

\subsection{Big Data Mining in C-RANs}

In the context of mobile networks, the big data, described by volume, variety, velocity, and value, includes subscriber-level, cell-level, core-network-level,
and other level data, which can facilitate the network towards a more proactive one \cite{son}.
Owing to the fast development of big data mining techniques and the powerful computing capability of the BBU pool,
it is feasible to utilize big data technology to extract interesting patterns or knowledge to enhance the self-organizing capabilities in C-RANs.
For example, by mining the historical content request data, users' interests in watching videos can be predicted, which helps edge devices
cache videos. However, there are still some technical challenges related to the big data mining in C-RANs.
For example, the transmission of the large volume of data collected by edge devices will put a heavy burden on the fronthaul.
Furthermore, the computing of sparse, uncertain and incomplete data is a big problem, which requires advanced data mining algorithms \cite{t-6}.

\subsection{Social-aware D2D in C-RANs}

D2D communication underlaying cellular system has been widely researched due to its superiority in improving SE and EE.
To take full advantages of D2D, several challenges should be handled, including peer discovery, mode selection,
resource allocation, and interference management \cite{social}. Recently, some works tackle
these challenges by leveraging the characteristics of social networks, including social tie \cite{tie} and social community \cite{community}.
As a result, the social-aware D2D is proposed.
For the C-RAN involving social-aware D2D, much data capacity can be offloaded from fronthaul links, alleviating the fronthaul constraints and shorten
the transmission latency.

Generally speaking, social-aware D2D in a cellular network can be implemented both with and without the assistance of
a BS \cite{t-8}. The social-aware D2D with the assistance of
a BS requires the BS to deliver control signalling, which is hard to implement in C-RANs since
RRHs are mainly deployed to provide high capacity in special zones \cite{HCRAN}. Hence, the social-aware D2D without the assistance of
a BS is preferred in C-RANs. In addition, the selection between D2D mode and C-RAN mode is critical to achieve high QoS.
Furthermore, when the D2D communication operates in the licensed band, the interference between D2D users and
RRH users can be severe, and meanwhile there can also exist mutual interference among D2D users.
Therefore, how to suppress those interference should be addressed.

\subsection{CR in C-RANs}

Facing with the ever increasing capacity demands,
the spectrum resource is becoming scarce under the conventional static spectrum allocation policy.
In contrast, due to the sporadical usage, a significant amount of assigned spectrum is not fully utilized \cite{t-10}.
A promising solution to the spectrum scarcity problem is to use the CR technology, which enables secondary users opportunistically communicate
over spectrum holes or share the spectrum with licensed users in an under or overlay manner \cite{t-11}.

For C-RANs, although centralized LSCP in the BBU pool can enhance SE,
CR is preferred to further improve the spectrum utilization rate. With the cognitive capability, RRHs can interact
with radio environment, and find temporally unused spectrum assigned to C-RANs.
Together with the CA technique, the operating bandwidth of RRHs can be greatly enlarged,
leading to higher data rates. While, the challenges mainly lie in the implementation complexity and the cost.
For example, to achieve spectrum sensing function, the complexity of transceiver is relatively high \cite{t-12}.
Moreover, to identify the vacant spectrum, conventionally, the secondary users need to continuously
sense the radio spectrum, and hence their transceivers should always stay in the active state \cite{t-13},
which results in consuming almost the same amount of power as the transmit state. In the future, the structure
of using CR in C-RANs should be further researched to reduce the implementation complexity and improve EE.

\subsection{SDN with C-RANs}

SDN decouples the control plane and data
plane, which provides a centralized controller
and network programmability \cite{sdn}.
By applying SDN to C-RANs, numerous benefits can be achieved.
For example, the software defined fronthaul proposed in \cite{soft} can achieve
flexible mapping between BBUs and RRHs, which adapts the network to traffic volume and user mobility.
Specifically, for a certain region with high mobility users,
multiple RRHs are logically mapped to a BBU by SDF to realize DAS, reducing handoffs and providing diversity gain.
When the region is with static users, SDF will configure one-to-one mapping to deliver high throughput.
In \cite{empowered}, a BBU pool is built in the cloud based on SDN and network function virtualization concepts, in which
RRHs are served by the BBU virtual instances.
Similar to the SDN in wired networks, two controllers are defined: infrastructure manager and service manager.
The infrastructure manager is responsible for pooling hardware resources and offering them as service slices
to the service manager, which consumes service slices according to the requirement of a BBU virtual instance.
By doing so, hardware resources in the BBU pool can be utilized more efficiently, reducing the cost of purchasing new hardware.

However, there are still some emerging challenges that should be addressed for the application of SDN to C-RANs.
Firstly, if only one central controller is in charge
of the whole C-RAN, the whole C-RAN may collapse in case that this controller fails.
Secondly, the controller placement should be optimized especially for cases with multiple controllers and the large-scale C-RAN,
since the placement has a significant impact on the processing latency and other network performance metrics.
Thirdly, limited by the service capability of the SDN controller, operators need to face with the scalability issues in C-RANs.

\subsection{Physical Layer Security in C-RANs}

Due to the broadcasting nature of wireless channels, the network is vulnerable to various newly arising attacks like the attack from eavesdroppers.
Meanwhile, considering that encryption often needs considerable computational resources and communication overheads,
in this context, physical layer security has been an attractive topic recently, which
exploits physical characteristics of wireless channels to achieve perfect secrecy against eavesdropping in an information-theoretic sense \cite{secure}.
In \cite{secure1}, opportunistic relaying is proposed to reduce the secrecy outage floor of cognitive communications. The security-reliability trade-off of relay-selection is considered in \cite{secure2}, which shows significant improvement compared to the classic direct transmission. Recognized as one of the future wireless networks, the C-RAN should ensure the communication security as well. To deal with this challenge, RRHs can be used to improve the security performance. Specifically,
in the downlink C-RANs, a UE is typically served by a cluster of RRHs, and some RRHs can act as jammers to
generate artificial noise to interfere with the eavesdroppers. However, some other challenges may arise, and one of them is the selection of jammers among the
available RRHs. Furthermore, how to design a novel architecture to efficiently support physical layer security in C-RANs is still not straightforward.

\subsection{Testbed and Trial Test}

In \cite{s1}, the notable achievements of C-RANs gained by China Mobile in trial tests and testbed development are elaborated.
As one of the biggest advocates for C-RANs, China Mobile has carried out
many trial tests in more than 10 cities in China.
First, by the TD-LTE based C-RAN field trials, the single fiber bi-direction and CPRI compression solution for the fiber resource saving is shown
to have little negative impact on the C-RAN performances.
Second, through the trial within China Mobile's TD-LTE
commercial networks in a dense urban area, it is found that all
performance metrics like throughput and handover successful rate are almost the same with or without
wavelength-division multiplexing which can carry dozens of carriers
on a single fiber. Third, based on the WDM test,
the gain of uplink LSCP is demonstrated,
reaching 40-100\% in the weak coverage area where reference signal received power is lower than -95 dBm.
Meanwhile, to validate the technical and business feasibility of C-RANs,
a GPP based C-RAN testbed has been developed in China Mobile, which can simultaneously support TD-LTE, FDD-LTE and GSM and
deliver a similar performance to the traditional DSP/FPGA based systems. To demonstrate the efficiency and achieving gains of C-RANs in practice, the large-size testbed and the corresponding trial test for C-RANs are anticipated.

\section{Conclusion}

This paper outlines and surveys the state-of-the-art system architecture, key technologies, and open issues in C-RANs. With the goal of understanding further intricacies of the key technologies, we have broadly divided the body of knowledge into fronthaul compression, large-scale collaborative processing, channel estimation, and cooperative radio resource allocation. Within each of these aspects, we have surveyed the diverse problems and the corresponding solutions that have been proposed for enabling C-RANs to be configured and operated in a high SE and EE performance manner.

Nevertheless, given the relative infancy of the field, there are still quite a number of outstanding problems that need further investigation from the perspective of proposing key techniques and advanced solutions. Given the extensiveness of the research areas, it is also concluded that more rigorous investigations are required with greater attention to be focused on transforming the well established fully-centralized C-RAN architecture into partially-centralized paradigm with edge cloud computing technology, in which edge cache and social-aware device-to-device are two key components. Furthermore, with the introduction of the advanced big data mining, cognitive radio, software defined network, and physical layer security, the availability of varied degrees of freedom along with the associated constraints further beckon the design and validation of the original models in the context of C-RANs. The progress of trial tests and test bed development of C-RANs can be anticipated to be promoted in the future, which makes C-RANs commercial rollout as early as possible.

\begin{IEEEbiography}[{\includegraphics[height=1.25in]{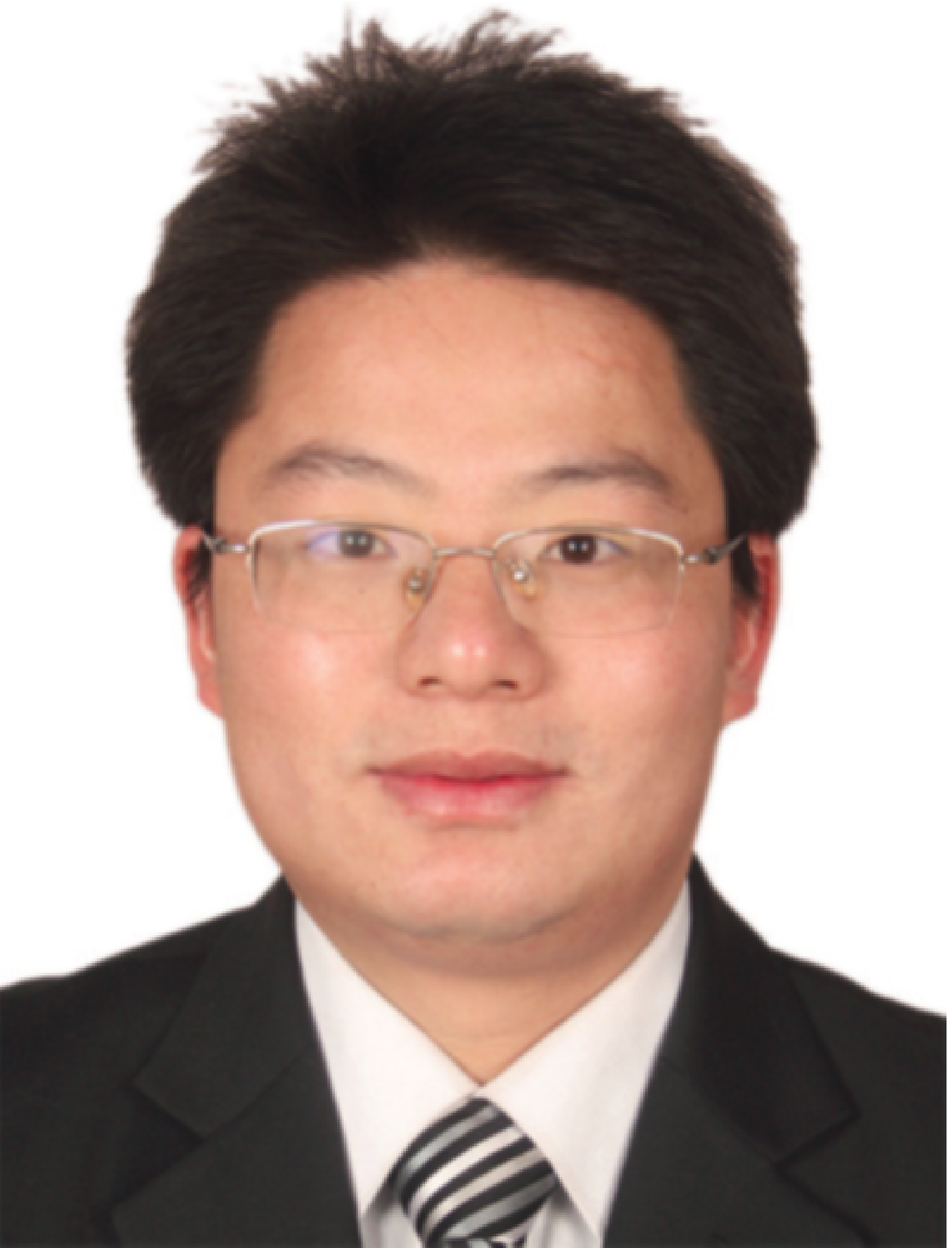}}]{Mugen Peng} (M'05--SM'11) received the B.E. degree in electronics engineering from the Nanjing University of Posts and Telecommunications, Nanjing, China, in 2000, and the Ph.D. degree in communication and information systems from the Beijing University of Posts and Telecommunications (BUPT), Beijing, China, in 2005. Afterward, he joined BUPT, where he has been a Full Professor with the School of Information and Communication Engineering since 2012. In 2014, he was an Academic Visiting Fellow with Princeton University, Princeton, NJ, USA. He leads a Research Group focusing on wireless transmission and networking technologies with the Key Laboratory of Universal Wireless Communications (Ministry of Education), BUPT. His main research areas include wireless communication theory, radio signal processing, and convex optimizations, with a particular interests in cooperative communication, self-organization networking, heterogeneous networking, cloud communication, and internet of things. He has authored/coauthored over 60 refereed IEEE journal papers and over 200 conference proceeding papers.

Dr. Peng was a recipient of the 2014 IEEE ComSoc AP Outstanding Young Researcher Award, and the best paper award in IEEE WCNC 2015, WASA 2015, GameNets 2014, IEEE CIT 2014, ICCTA 2011, IC-BNMT 2010, and IET CCWMC 2009. He received the First Grade Award of the Technological Invention Award in the Ministry of Education of China for the hierarchical cooperative communication theory and technologies, and the First Grade Award of Technological Invention Award from the China Institute of Communications for the contributions to the self-organizing networking technology in heterogeneous networks. He is on the Editorial/Associate Editorial Board of the \emph{IEEE Communications Magazine}, \emph{IEEE Access}, \emph{IET Communications},
\emph{International Journal of Antennas and Propagation (IJAP)},
and \emph{China Communications}. He has been the guest leading editor
for the special issues in the \emph{IEEE Wireless Communications}, \emph{IEEE Access}, and \emph{IET Communications}. He is the Fellow Of the Institution of Engineering and Technology (IET).
\end{IEEEbiography}

\begin{IEEEbiography}{Yaohua Sun}
received his bachelor degree in Telecommunications Engineering with Management from Beijing University of Posts and Telecommunications (BUPT) in 2014 with first class honors, and also got BUPT Excellent Project Award. During the undergraduate study, he won the national scholarship and the Scholarship of Tang Jun and Sun Chunlan. He is currently a Ph.D. student in the Key Laboratory of Universal Wireless Communications (Ministry of Education) at BUPT. His main research interests include coalition game, Stackelberg game, contract theory, matching theory, and their applications to resource management in fog computing based radio access networks.
\end{IEEEbiography}

\begin{IEEEbiography}{Xuelong Li}(M'02-SM'07-F'12) is a full professor with the Center for OPTical IMagery Analysis and Learning (OPTIMAL), State Key Laboratory of Transient Optics and Photonics, Xi'an Institute of Optics and Precision Mechanics, Chinese Academy of Sciences, Xi'an 710119, Shaanxi, P. R. China.
\end{IEEEbiography}

\begin{IEEEbiography}{Zhendong Mao}
received the B.S. degree in applied physics from the Beijing
University of Posts \text{\&} Communications (BUPT), China, in 2013.
He is currently pursuing the Ph.D. degree at BUPT. His research
interests focus on the channel estimation and training design in cloud radio
access networks (C-RANs) and fog radio access networks.
\end{IEEEbiography}

\begin{IEEEbiography}{Chonggang Wang}
(SM'09) received the B.S. degree from Northwestern Polytechnical
University (NPU), Xi'an, China, in 1996, the M.S. degree from the
University of Electronic Science and Technology of China (UESTC),
Chengdu, China, in 1999, and the Ph.D. degree from the Beijing
University of Posts and Telecommunications (BUPT), Beijing, China,
in 2002. He is currently a Member Technical Staff with the
Innovation Lab, InterDigital Communications, King of Prussia, PA,
USA, with a focus on Internet of Things (IoT) R\&D activities
including technology development and standardization. His current
research interest includes IoT, machine-to-machine (M2M)
communications, mobile and cloud computing, and big data management
and analytics. Dr. Wang was a corecipient of the National Award for
Science and Technology Achievement in Telecommunications in 2004 on
IP Quality of Service (QoS) from the China Institute of
Communications. He is the founding Editor-in-Chief of IEEE Internet
of Things Journal, an Advisory Board Member of The Institute-IEEE
(2015-2017), and on the editorial board for several journals
including IEEE Transactions on Big Data and IEEE Access. He has been
selected as an IEEE ComSoc Distinguished Lecturer (2015-2016).
\end{IEEEbiography}

\end{document}